\documentclass[conference]{IEEEtran}
\usepackage{}
\usepackage{graphicx}
\usepackage{amsfonts}
\usepackage{amsmath}
\usepackage{times}
\usepackage{fancyhdr}
\usepackage{algo}
\usepackage{wrapfig}
\usepackage{graphicx}
\usepackage{caption}
\usepackage{subcaption}
\usepackage{epsfig}
\usepackage{epstopdf}
\usepackage[linesnumbered]{algorithm2e}
\usepackage[numbers]{natbib}
\usepackage{multicol}
\usepackage{epsfig} %

\DeclareMathOperator{\sign}{sign}

\newcommand{\va}{{\bf a}}

\newcommand{\vx}{{\bf x}}
\newcommand{\vy}{{\bf y}}
\newcommand{\vz}{{\bf z}}

\newcommand{\vu}{{\bf u}}

\newcommand{\vI}{{\bf I}}

\newcommand{\vW}{{\bf W}}

\newcommand{\vK}{{\bf K}}

\newcommand{\vQ}{{\bf Q}}
\newcommand{\vB}{{\bf B}}

\newcommand{\vX}{{\bf X}}

\newcommand{\vP}{{\bf P}}
\newcommand{\vA}{{\bf A}}

\newcommand{\VSigma}{{\mbox{\boldmath$\Sigma$}}}
\newcommand{\valpha}{{\mbox{\boldmath$\alpha$}}}

\newcommand{\vcalP}{{\mbox{\boldmath$\cal{P}$}}}

\newcommand{\argmax}{\operatornamewithlimits{argmax}}

\newcommand{\rd}{{\mathrm d}}
\newcommand{\rp}{{\mathrm p}}

\newcommand{\rT}{{\mathrm{T}}}

\begin{document}

\title{\LARGE \bf Nonparametric Infinite Horizon Kullback-Leibler Stochastic Control}


\author{\IEEEauthorblockN{Yunpeng Pan}
\IEEEauthorblockA{Daniel Guggenheim School of Aerospace Engineering\\
Georgia Institute of Technology\\
Atlanta, Georgia 30332--0250\\
Email:ypan37@gatech.edu}
\and
\IEEEauthorblockN{Evangelos A. Theodorou}
\IEEEauthorblockA{Daniel Guggenheim School of Aerospace Engineering\\
Georgia Institute of Technology\\
Atlanta, Georgia 30332--0250\\
Email: etheodorou3@mail.gatech.edu}}



%


\maketitle

\begin{abstract}
We present two nonparametric approaches to Kullback-Leibler (KL) control, or linearly-solvable Markov decision problem (LMDP) based on Gaussian processes (GP) and Nystr\"{o}m approximation. Compared to recently developed parametric methods, the proposed data-driven frameworks feature accurate function approximation and efficient on-line operations. Theoretically, we derive the mathematical connection of KL control based on dynamic programming  with earlier work in control theory  which  relies on  information theoretic dualities for  the infinite time horizon case. Algorithmically, we give explicit optimal control policies in nonparametric forms, and propose on-line update schemes with budgeted computational costs. Numerical results demonstrate the effectiveness and usefulness of the proposed frameworks.
\end{abstract}

\section{Introduction}
Stochastic optimal control plays one of the key roles in motor control of complex nonlinear systems. Recently, Kullback Leibler (KL) control, or linearly-solvable Markov decision problem (LMDP) has demonstrated remarkable applicability to robotic control and planning problems \citep{kinjo2013evaluation}. In \citep{todorov2006linearly}\citep{todorov2009efficient}, the Bellman principle of optimality was applied for discrete time optimal control problems in which the control cost is formulated as the KL divergence between the controlled and uncontrolled dynamics. The resulting framework applies  to a large class of control problems which include finite, infinite horizon, exponentially discounted and first exit \citep{todorov2009efficient}.

One of the most challenging tasks in KL control or LMDP is the approximation of desirability function defined in continuous state space. Recently, parametric approaches have been developed and implemented in real robotic systems \cite{todorov2009eigenfunction}\citep{kinjo2013evaluation}. Although the linearly-solvable formulation has shown superior efficiency compared to policy and value iteration \citep{todorov2006linearly}\cite{todorov2009eigenfunction}, there are still  major limitations with these parametric methods. Despite the  guaranteed convergence, the parameters of the basis functions used to approximate the desirability function  may converge to the wrong solution depending on the control problem under consideration\cite{todorov2009eigenfunction}. This behavior results in inaccurate approximation of the underlying desirability function.


In this paper, we provide a unified view of KL  control for the infinite time horizon case.  This unified view brings together earlier work in control theory \cite{Fleming:1995} with more recent work in machine learning and robotics \citep{kinjo2013evaluation,todorov2006linearly,todorov2009efficient}.   In particular, we show  two alternative derivations of KL control which rely on  the dynamic programming principle and the information theoretic dualities between  free energy and relative entropy, respectively. We propose two nonparametric  frameworks  for infinite horizon KL control.  The first framework is based on Gaussian processes (GP) \citep{williams2006gaussian}, which is a Bayesian modeling approach with data-driven, generative models. The second framework is based on Nystr\"{o}m approximation, which is considered as a sampling-based low-rank matrices approximation method and is originated from the numerical solver of integral equations \cite{baker1977numerical}. We employ both GP and Nystr\"{o}m method to approximate desirability functions associated with KL control.  We show the nonparametric forms of the corresponding optimal control policies and we present  efficient on-line update schemes   to improve applicability.

%
 
The rest of the paper is organized as follows: In Section II and III, we derive KL control based on both dynamic programming and information theoretic dualities, and show their connections for infinite time horizon case. Section IV and V introduce the proposed nonparametric frameworks for KL control based on GP and Nystr\"{o}m approximation. Numerical results are provided and discussed in Section VI. Finally Section VII concludes this paper.

\section{Infinite Horizon Kullback Leibler Control based on Dynamic Programming}

\subsection{Infinite Horizon Stochastic Optimal Control}
\label{sec:ContinousStochasticControl}

We consider the stochastic optimal control problem with state $\vx\in\mathbb{R}^D$ and control $\vu\in\mathbb{R}^m$ of the following form:
\begin{equation}
  v(\vx)  =\min_{\vu} \lim_{T \to \infty} \frac{1}{T} \mathbb{E} \bigg[  \int_{0}^{T} {{\cal{L}}(\vx(t), \pi(\vx(t))\rd t} \bigg],\nonumber
\end{equation}
subject to the dynamics: $\rd \vx = \valpha (\vx)\rd t + \vB(\vx) (\vu \rd t + \sigma \rd\omega)$,
and the cost rate function: ${\cal{L}}(\vx(t), \vu(t)) = q(\vx) + \frac{1}{2 \sigma^{2}} || \vu ||^{2}$.
The function $ q(\vx) \geq 0 $  and the drift $ \valpha(\vx) $ and diffusion $\vB(\vx) $. Let $v(\vx)$ be the value function and $v_x(\vx)$ its gradient. The optimal control control has the following form:
\begin{equation}\label{optimalcontrol}
\vu = -\sigma^2\vB(\vx)^{\rT}v_x(\vx).
\end{equation}
The value function $ v(\vx) $ satisfies the Hamilton-Jacobi-Bellman (HJB)
equation
\begin{equation} \label{HJB1}
  c = q(\vx) + {\cal{D}} [ v] (\vx) - \frac{1}{2} v_{\vx}^{\rT}(\vx) \Sigma(\vx)  v_{\vx}(\vx),
\end{equation}
 where the linear Differential operator $  {\cal{D}}$ is defined as
 \begin{equation}
 {\cal{D}} [v] (\vx)  =  \valpha(\vx)^{\rT}  v_{\vx} + \frac{1}{2} \text{tr}  \bigg( \Sigma(\vx)  v_{\vx\vx}  \bigg),
 \end{equation}
and $ \Sigma $  is the noise covariance matrix: $ \VSigma(\vx) =  \sigma^{2} \vB(\vx) \vB(\vx)^{\rT}$. For the case of infinite horizon average-cost-per step, $  c>0  $ is the unknown average cost-per-step and $ v $ is the differential operator 
 cost-to-go.  Moreover, for the case of first-exit problems, $c = 0 $  and $  v $ is the actual cost-to-go. The HJB equation takes a linear form under the exponential transformation $ z(\vx) =  \exp(-  v (\vx))$. 
 By exponentiating $ v(\vx) $ we get
 \begin{equation}\label{LinearBellman0}
    (q(\vx) - c) z(\vx) = {\cal{D}}[z](\vx) .
 \end{equation} 
 
\subsection{Discrete time formulation}
\label{sec:LK_DiscreteFormulations}

In the discrete time case, the stochastic dynamics are discretized and therefore $ \vx(k) $ in discrete time corresponds to 
$ \vx(k \rd t) $.  In addition the controller specifies the transition probability  $ \pi(\vy| \vx) $. Therefore in discrete time we will have: $\vx(k+1) \sim \pi(\cdot|\vx(k))$. The cost rate is now formulated as
   \begin{equation}
     {\cal{L}}(\vx,\pi(\cdot|\vx)) = q(\vx)dt +\mathbb{KL}(\pi(\cdot| \vx) || \rp(\cdot|\vx)),
   \end{equation}
with the term  $ \pi(\cdot| \vx)$ denoting the transition probability under the control dynamics and   $ \rp(\cdot|\vx) $ is the transition probability under the uncontrolled dynamics. 
The distribution under the optimal control law is
\begin{equation}
   \pi^{*}(\vy|\vx)= \frac{\rp(\vy|\vx) z(\vy)}{{\cal{G}} [z](\vx)}.
\end{equation}
The term $  {\cal{G}}  $ is a linear integral operator defined as ${\cal{G}} [z] (\vx) =  \int \rp(\vy|\vx) z(\vy) \rd \vy$. The minimized Bellman equation can now be exponentiated and expressed in terms of $  z $  as follows
\begin{equation}\label{LinearBellman1}
\exp\bigg( \rd t q(\vx) - \rd t c  \bigg) z(\vx) =  {\cal{G}} [z] (\vx) .
\end{equation}
It has been shown that  the equation above has a unique positive solution $ z(\vx) $ that corresponds to the largest eigenvalue \cite{todorov2006linearly} $ \lambda  = \exp(-dtc) $. For the case of  discrete-state Markov Decision Process (MDP) we define a set of states  $ \{\vx_{n}\} $.  In this case equation  (\ref{LinearBellman1}) can be rewritten as
\begin{equation}\label{Discrete-MDP}
 \lambda z = \vQ \vP z,
\end{equation}
with  $  z(\vx_{n})  $ the desirability function at every states  $ \vx_{n} $ in the set, $\vQ$ a diagonal matrix of elements $ \exp(- \rd t q(\vx_{n}))$ and $\vP $ the passive transition probability matrix.  This eigenvalue problem can be solved by various methods such as power iteration.

\subsection{Relationship between Continuous and Discrete Case}

  To make the connection with the continuous case we represent the passive dynamics  $ p(\vy|\vx) $ as 
\begin{equation}\label{PassiveDynamicsProbability}
   \rp(\vy|\vx) = {\cal{N}} (\vy; \vx+ \rd t \va(\vx), \rd t \VSigma(\vx)).
\end{equation}
Consider the generator of a stochastic process:
\begin{equation}
\lim_{\rd t \to 0} \frac{\mathbb{E}[ z(\vy) | \vy(0) = \vx] - z(\vx)}{\rd t} = {\cal{D}}[z](\vx) .
\end{equation}
 Since $ {\cal{G}} [z] (\vx)   = \mathbb{E}[ z(\vy) | \vy(0) = \vx]$  we will have
 \begin{equation}\label{LinearBellman_2}
       {\cal{G}} [z] (\vx) = z(\vx)  + \rd t {\cal{D}}[z](\vx)   + o(\rd t^{2}).
 \end{equation} 
Substitute (\ref{LinearBellman_2}) into (\ref{LinearBellman1}) results in (\ref{LinearBellman0}).

\section{Kullback Leibler  Control based on Information theoretic Dualities}
   \label{sec_dualities}

  The work in \cite{TheodorouCDC2012} have shown the mathematical links  between 1) the  information theoretic point of view of stochastic optimal control theory as presented within the control theory community,  and  2) the path integral and Kullback Leibler  formulations for  finite horizon stochastic control as presented within  machine learning and statistical physics communities.  In this section we will show the corresponding connections for the case of infinite horizon stochastic optimal control problems. 

 To do so,  below we provide the definition of free energy and relative entropy and review the Legendre transformation. In particular, Let  $( \Omega, {\cal{F}}) $  be a measurable space, where  $ \Omega $   denotes the sample space and   $ {\cal{F}}$ denotes a $ \sigma$-algebra, and  let  $ \vcalP(\Omega) $ define a probability  measure on the $ \sigma$-algebra $ \cal{F}$.
For our analysis we consider the following definitions.

  \textbf{Definition 1:}  Let  $   \mathbb{P} \in  \vcalP(\Omega)$ and let  the function   ${\cal{J}}(\vx): \Omega \to  \Re$ be a measurable function. Then
  \begin{equation}
{\cal{E}} \bigg( {\cal{J}}(\vx)  \bigg)  = \log_e   \int \exp(\rho\cal{J}(\vx))   \rd \mathbb{P},
  \end{equation}
      is called free energy of  $ {\cal{J}}(\vx)  $    with respect to $   \mathbb{P} $.

     \textbf{Definition 2:} Let  $   \mathbb{P} \in  \vcalP(\cal{Z})$  and  $   \mathbb{Q} \in  \vcalP(\cal{Z})$, the relative entropy of $   \mathbb{P} $  with respect to    $   \mathbb{Q} $ is defined as
   \[ \mathbb{KL}(\left( \mathbb{Q} ||  \mathbb{P}  \right) = \left\{
\begin{array}{l l}
  \int \log_e \frac{\rd\mathbb{Q}}{\rd\mathbb{P}} \rd\mathbb{Q}
 & \quad \mbox{if $\mathbb{Q}<<\mathbb{P} $, $ \log_e \frac{\rd\mathbb{Q}}{\rd\mathbb{P}} \in L^{1}$  }  \\
  +\infty & \quad \mbox{otherwise}\\ \end{array} \right. \]
where ``$<<$'' denotes absolute continuity of $\mathbb{Q}$ with respect to $\mathbb{P}$ and
$\mathcal{L}_1$  denotes the space of Lebesgue measurable functions on  $ [0, \infty)$. We say that $ \mathbb{Q} $ is \textit{absolutely continuous} with respect to $ \mathbb{P} $ and we write  $\mathbb{Q}<<\mathbb{P} $ if  $ \mathbb{P}(H)  = 0 \Rightarrow \mathbb{Q}(H) = 0, ~ \forall H \in \cal{F}$.    We will also consider  the   function
    \begin{equation}
      \xi(\vx,T) =\frac{1}{\rho}   {\cal{E}}  \bigg( {\cal{J}}(\vx,T)  \bigg)  =   \frac{1}{\rho} \log_e  \mathbb{E}_{\mathbb{P}}\bigg[ \exp{\left( \rho {\cal{J}}(\vx,T) \right)}   \bigg] \nonumber,
    \end{equation}
    with  ${\cal{J}}(\vx,T) $ $=  \phi(\vx_{t_{N}}) +  \int_{0}^{T} q(\vx) \rd t$  is the state depended cost.   The  objective function above  takes  the form       $   \xi(\vx) = \mathbb{E}_{\mathbb{P}}\left(  \cal{J} \right)  +\frac{\rho}{2}   {{\mathbb{VAR}}} \left( \cal{J}    \right) $ as $ \rho \to 0  $.  \footnote{For small $ \rho $  the  cost is a function of the mean the variance.  When $ \rho>0 $ the cost function is risk sensitive while for $ \rho<0 $   is  risk seeking.} To derive the basic relationship between free energy and relative entropy we express the expectation $    \mathbb{E}_{\mathbb{P}}$ taken under the measure $ \mathbb{P}$ as a function of the expectation  $ \mathbb{E}_{\mathbb{Q}}  $ taken under the probability measure $ \mathbb{Q}$.   More precisely will have
    \begin{align*}
   \mathbb{E}_{\mathbb{P}}\bigg[ \exp{\left( \rho {\cal{J}}(\vx,T) \right)}   \bigg]          & =  \int  \exp{\left( \rho {\cal{J}}(\vx,T) \right)}  \frac{\rd\mathbb{P}}{\rd\mathbb{Q}}  \rd \mathbb{Q}.
          \end{align*}
          By taking the logarithm of both sides of the equations above and making use of the Jensen's inequality we will have
           \begin{align*}
          \log_e  \mathbb{E}_{\mathbb{P}} \bigg[ \exp{\left( \rho \cal{J}(\vx) \right)}   \bigg] &  \geq   \int \log_e  \bigg(   \exp{\left( \rho \cal{J}(\vx) \right)}  \frac{\rd\mathbb{P}}{\rd\mathbb{Q}}  \bigg) \rd \mathbb{Q} \\
  & =  \int  \rho { \cal{J}}(\vx) \rd \mathbb{Q}  -\mathbb{KL}\left( \mathbb{Q} ||  \mathbb{P}  \right).
    \end{align*}
      We multiply the inequality above with $ \frac{1}{\rho} $  for case of $ \rho <0 $   or $ \rho = - | \rho| $  and thus  we  have
    \begin{equation}\label{Legendre_Transformation}\boxed{
      \xi(\vx) = - {\frac{1}{| \rho| }    {\cal{E}}  \left( {\cal{J}}(\vx) \right)  \leq   \mathbb{E}_{\mathbb{Q}} \left( \cal{J}(\vx)  \right)} + \frac{1}{|\rho|} \mathbb{KL}\left( \mathbb{Q} ||  \mathbb{P} \right)}.
    \end{equation}
The inequality above gives us the duality relationship between relative  entropy and  free energy. Essentially one could define the following minimization problem
    \begin{equation} \label{Dual1}
     -{\frac{1}{|\rho|}   {\cal{E}}  \left( {\cal{J}}(\vx)  \right) =  \inf  \bigg[ \mathbb{E}_{\mathbb{Q}} \left(\cal{J}(\vx)  \right)} + \frac{1}{|\rho|} \mathbb{KL}\left( \mathbb{Q} ||  \mathbb{P} \right) \bigg].
    \end{equation}
The infimum in  (\ref{Dual1}) is attained at  $   \mathbb{Q^{*}}$ given by
       \begin{equation}\label{OptimalDistribution1}
          \rd  \mathbb{Q}^{*}  = \frac{\exp{\left(- | \rho|  {\cal{J}}(\vx) \right) } \rd \mathbb{P}}{\int \exp{\left(- | \rho|  {\cal{J}}(\vx) \right)}\rd \mathbb{P}}.
       \end{equation}
   In the next section we follow the steps of  \cite{Fleming:1995}  to show how inequality (\ref{Dual1}) is transformed to a stochastic optimal control problem for the case of Markov diffusion processes.


    \subsection{Connections to Dynamic Programming}

    We start our analysis with the free energy term in the  Legendre transformation (\ref{Legendre_Transformation}). We follow the steps as in \cite{Fleming:1995} but for the case of the risk seeking version of the free energy. Since our analysis is on  infinite horizon stochastic control case we divide the free energy  term  with $ \frac{1}{T} $ and take the limit as $ T \to \infty $. In addition, to simplify our analysis we will assume $ \rho = 1$.      More precisely
   \begin{align} \label{Legendre_Transformation_Infinite}
    \mu &=-  \lim_{T \to \infty}  \frac{1}{T}  \log_e   \underbrace{\mathbb{E}_{\mathbb{P}}\bigg[ \exp{\left(-  {\cal{J}}(\vx,T) \right)}   \bigg] }_{\phi(\vx,T)}  \nonumber   \\
       & \leq \lim_{T \to \infty}  \frac{1}{T}\bigg[  \mathbb{E}_{\mathbb{Q}} \left( \cal{J}(\vx)  \right) + \mathbb{KL}\left( \mathbb{Q} ||  \mathbb{P} \right) \bigg]   .
           \end{align}
 The function  $ \phi(\vx,T)  $   is the parameterized, by the time horizon $ T $, desirability function. The expectations    $ \mathbb{E}_{\mathbb{P}} $ and $ \mathbb{E}_{\mathbb{Q}} $ are taken over state trajectories generated with forward sampling of the dynamics
 \begin{align}
   \rd \vx &= \valpha (\vx)\rd t + \vB(\vx) \sigma \rd\omega, \label{UncontrolledDynamics} \\
   \rd \vx &= \valpha (\vx)\rd t + \vB(\vx)  (\vu \rd t + \sigma \rd\omega). \label{ControlledDynamics}
 \end{align}
   The desirability function satisfies the PDE that follows
  \begin{equation}\label{BackwardChapmanKolmogorov}
    \frac{\partial \phi(\vx,T) }{\partial T } = {\cal{D}}[\phi](\vx)  -  q(\vx) \phi(\vx,T),
  \end{equation}
which is a form of the Feynman-Kac formula \cite{Fleming:1995}. It is different from the forms used so far in the finite horizon stochastic optimal control case \cite{TheodorouCDC2012}. Next we formally separate variables as in \cite{Fleming:1995}, namely 
   \begin{equation}\label{VariableSeparation}
       \phi(\vx,T) \sim \exp(- \mu T) z(\vx).
   \end{equation}
Substitute back into (\ref{BackwardChapmanKolmogorov}) results in 
\begin{equation}\label{PDE}
- \mu z(\vx)  = {\cal{D}}[z](\vx)  - q(\vx)  z(\vx).
\end{equation}
Which can be further written as 
  \begin{equation}\label{LinearBellman2}
       ( q(\vx)     -  \mu ) z(\vx)  = {\cal{D}}[z](\vx) .
 \end{equation}
For $ \mu  = c $ the equation above is exactly the same a (\ref{LinearBellman0}). Up to this point, we have shown the  equivalence  with the approach in Section \ref{sec:LK_DiscreteFormulations}  and \ref{sec:ContinousStochasticControl}. Next we go one step further by deriving the HJB equation. More precisely,   under  the  exponential   transformation   of $ z(\vx) = \exp(-v(\vx)) $ the equation  (\ref{VariableSeparation}) heuristically takes the form  
\begin{align}\label{heuristic}
- \log_e  \phi(\vx,T)   \sim \mu T + v(\vx).
\end{align}
To show the validity/motivation of (\ref{heuristic}), multiply both side with $\frac{1}{T}$ and take limit as $T\to\infty$. We have $\lim_{T \to \infty}-\frac{1}{T}\log_e \phi(\vx,T)=\mu$ which is the same as (\ref{Legendre_Transformation_Infinite}).
In addition, substitute the exponential transformation $ z(\vx) = \exp(-  v(\vx) ) $  into   ({\ref{PDE}) and taking into account that $ z_{\vx \vx} = -  z v_{\vx}  v_{\vx}^{T} +  z  v_{\vx\vx}$  and  $ z_{\vx}= - z v_{\vx} $  results in:
\begin{align*}
{\cal{D}} [z] (\vx) =-  z \valpha(\vx)^{\rT}   v_{\vx}  +  \frac{1}{2}  z   v_{\vx}^{\rT} \vB  \vB^{\rT}    v_{\vx}   -  \frac{1}{2}    z\text{tr} \bigg(    v_{\vx \vx}\vB \vB^{T} \bigg) .
\end{align*}
Substitution of the operator above into  (\ref{PDE}) results in:
   \begin{equation}
   \mu  =  \valpha(\vx)^{\rT} v_{\vx}  -  \frac{1}{2}    v_{\vx}^{\rT} \vB \sigma^{2} \vB^{\rT}   v_{\vx}  +  \frac{1}{2}   \text{tr} \bigg( \Sigma(\vx)  v_{\vx \vx}  \bigg)   +   q(\vx). \nonumber
 \end{equation}
The above equation is the Hamilton-Jacobi-Bellman PDE for the case of infinite time horizon stochastic control and is exactly the same with (\ref{HJB1}). As we can see   $ \mu $ and $ v(\vx) $ satisfy the HJB equation. The term   $ \mu  $  is the optimal     expected cost per unit time and $ v(\vx) $ is the associated cost potential function.  The optimal control is formulated as  $ \vu(\vx) = - \sigma^{2} \vB^{\rT}  v_{\vx} = \sigma^{2} \vB^{\rT} \frac{z_{\vx}}{z} $.  Finally we make use of  the derivative   $  \frac{\rd \mathbb{P}}{\rd \mathbb{Q}} $, that is the Radon-Nikodym derivative  \cite{Kushner1991}  for the stochastic dynamics in (\ref{UncontrolledDynamics}) and  (\ref{ControlledDynamics}) to find an expression for the Legendre transformation in  (\ref{Legendre_Transformation_Infinite_Horizon}). This expression  completes the connection with stochastic optimal control formulation. More precisely we  have:
 \begin{equation}
    \mathbb{KL}\left( \mathbb{Q} ||  \mathbb{P} \right)  = \mathbb{E}_{\mathbb{Q}} \bigg(\log_e  \frac{\rd \mathbb{P}}{\rd \mathbb{Q}}  \bigg) =  \mathbb{E}_{\mathbb{Q}}\bigg( \int_{0}^{T}  \frac{1}{2 \sigma^{2}}||\vu||^{2}   \rd t\bigg) .
 \end{equation}
  Based on this last result  the Legendre transformation  takes the form
   \begin{align} \label{Legendre_Transformation_Infinite_Horizon}
    \mu &=-  \lim_{T \to \infty}  \frac{1}{T}  \log_e  \mathbb{E}_{\mathbb{P}}\bigg[ \exp{\left(-  {\cal{J}}(\vx,T) \right)}   \bigg] \nonumber   \\
       & \underbrace{ \leq \lim_{T \to \infty}  \frac{1}{T}\bigg[  \mathbb{E}_{\mathbb{Q}} \left( \int_{0}^{T} \left(  q(\vx) + \frac{1}{2\sigma^{2}} || \vu||^{2}  \right)\rd t \right) \bigg]}_{\textbf{Stochastic Optimal Control Cost}}   .
           \end{align}
The left-hand side of the inequality is the control cost under minimization that is lower bounded by $ \mu $.   This last  form of Legendre transformation  completes the connection with stochastic optimal control.

\section{Gaussian Processes for KL Control}

The desirability $z(\vx)$ is a well-defined functional mapping of continuous, possibly high-dimensional inputs to scalar outputs. The goal is to infer the desirability given a newly observed state. This inference can be viewed as a nonlinear regression problem.  In this section, we introduce an on-line Gaussian process approach to KL control (GP-KL). In the rest of the section, consider that we are given a set of N observed states $\mathcal{X}=\{\vx_1,\vx_2,\ldots,\vx_N\}$ and corresponding desirability functions $\mathcal{Z}=\{z(\vx_{1}), z(\vx_{2}),...,z(\vx_{N})\}  $  where each state $\vx\in \mathbb{R}^{D}.$   We can define the state set as a vector $\vX=[\vx_1, \vx_2, \ldots, \vx_N]\in \mathbb{R}^{D\times N}$ and corredponding desirability $z(\vX) =[z(\vx_{1}), z(\vx_{2}),...,z(\vx_{N})]\in\mathbb{R}^{1\times N} $ .

\subsection{Desirability learning via Gaussian process regression} \label{gpr_desirability}
A GP is defined as a collection of random variables, any finite number subset of which have a joint Gaussian distribution. Given an input vector $\vx$, and corresponding output $z(\vx)$, a GP is completely defined by a mean function and a covariance function. The joint distribution of the observed output and an unknown output corresponding to a given test input $\vx^*$ can be written as
\begin{equation}\nonumber \small{
\rp\left( \begin{array}{c}
z(\vX) \\
z(\vx^*) \end{array} \right) \sim \mathcal {N} \bigg(0, \left[ \begin{array}{cc}
\vK(\vX,\vX)+\sigma_{n}^2\vI & \vK(\vX,\vx^*) \\
\vK(\vx^*,\vX) & \vK(\vx^*,\vx^*) \end{array} \right]\bigg) }.
\end{equation}
The covariance of this multivariate Gaussian distribution is defined via a kernel matrix $\vK(\vx_i,\vx_j)$. $\sigma_{n}$ represents zero-mean white noises.  In particular, in this paper we consider the Gaussian kernel, which is most widely used kernel function $\vK(\vx_i,\vx_j)=\sigma_s^2\exp(-\frac{1}{2} (\vx_i-\vx_j)^\rT\vW  (\vx_i-\vx_j)) + \sigma_n^2$,
with $\sigma_s,\sigma_n,\vW$ the hyper-parameters of the GP.  The kernel function can be interpreted as a similarity measure of random variables. More specifically, if inputs $\vx_i$ and $\vx_j$ are close to each other in the kernel space, their output $z(\vx_i)$ and $z(\vx_j)$ are highly correlated. The posterior distribution, which is also a Gaussian, can be obtained by constraining the joint distribution to contain the output $z(\vx^*)$ that are consistent with the observations. Therefore, the predictive distribution can be computed by conditioning the joint prior over the observed output as
\begin{align}
\rp\Big(z(\vx^*)\vert z(X),\vx^*,\vX\Big) \sim \mathcal {N}\Big(\bar{z}(\vx^*),\Sigma[z(\vx^*)]\Big), \nonumber
\end{align}
where the mean and variance are given by
\begin{equation}\label{gp_for_z}
\begin{split}
\bar{z}(\vx^*)=& \vK(\vx^*,\vX)(\vK(\vX,\vX)+\sigma_{n}^2\vI )^{-1}z(\vX), \\
\Sigma[z(\vx^*)]=&-\vK(\vx^*,\vX)(\vK(\vX,\vX)+\sigma_{n}^2\vI )^{-1}\vK(\vX,\vx^*)\\
 &+\vK(\vx^*,\vx^*).
\end{split}
\end{equation}
The kernel or hyper-parameter $\Theta=(\sigma_n,\sigma_s,\vW)$ can be learned by maximizing the log-likelihood of the training outputs given the inputs
\begin{equation}\label{hyp}
\Theta^*=\argmax_{\Theta}\bigg\{\log\bigg(\rp\Big(z(\vX)|\vX,\Theta\Big) \bigg)\bigg\}.
\end{equation}
The optimization problem can be solved using numerical methods such as conjugate gradient \cite{williams2006gaussian}.
 We use the mean of predictive distribution to infer the desirability function of a given newly observed state $\vx^*$.

\subsection{Optimal control policy based on GP}\label{compute_u}
Given the approximated desirability function $z(\vx^*)$, we can compute the optimal control as:
\begin{equation}\label{ux}
 \vu(x^*)=-\sigma^{2}\vB(\vx^*)^T v_{\vx^*}(\vx^*) =\sigma^{2}\vB(\vx^*)^T\frac{z_{\vx^*}(\vx^*)}{z(\vx^*)},
\end{equation}
where
$z_{\vx^*}(\vx^*)=[\frac{\partial z(\vx^*)}{\partial \vx^{(1)*}},\ldots,\frac{\partial z(\vx^*)}{\partial \vx^{(d)*}},\ldots,\frac{\partial z(\vx^*)}{\partial \vx^{(D)*}}]$. With each element
\begin{equation}
\frac{\partial z(\vx^*)}{\partial \vx^{(d)*}}=\frac{\partial \vK(\vx^*,\vX)}{\partial \vx^{(d)*}}(\vK(\vX,\vX)+\sigma_{n}^2\vI )^{-1}z(\vX),
\end{equation}
where $\frac{\partial \vK(\vx^*,\vX)}{\partial \vx^{(d)*}}$ can be evaluated analytically.
Due to the data-driven nature of the proposed GP-based approach, the optimal control policy can be computed without assumed parameterizations as in \citep{todorov2009eigenfunction}. In the next subsection, we will show how to further improve the on-line efficiency of the proposed framework.

\subsection{On-line update of optimal control policy}
One way of applying the control policy on-line is to incorporate every newly observed state to the state set $\mathcal{X}$ at each time step. For instance, let the   state space vector be  $\vX\in\mathbb{R}^{D\times N}$ at time t. At time step t+1, the newly observed state vector becomes
\begin{equation}
\vX_{t+1}=[\vX_{t}, \vx^*_{t}]\in \mathbb{R}^{D\times (N+1)}.
\end{equation}
However, as the observed state vector grows, the size of kernel matrices used for computing optimal control policy grows as well, such that $\vK(\vX_{t+1},\vX_{t+1})\in \mathbb{R}^{(N+1)\times (N+1)}$. In particular, when computing the inverse of kernel matrix $\vK(\vX_{t+1},\vX_{t+1})^{-1}$, the time complexity becomes $\mathcal {O} (N+1)^3$, which will increase cubically over time. Therefore, computing the optimal control policy on-line would become computationally intractable for the infinite horizon case. Now we introduce an on-line update scheme with fixed size of kernel matrices based on sparse Gaussian processes \citep{csato2002sparse}.

Our goal is to compute the optimal control policy without increasing the dimension of the kernel matrices. To do so, we introduce a scheme to delete one state from $\mathcal{X}$ when a newly observed state $\vx_t^*$ is incorporated at $t+1$. To determine whether or not to add/delete a state from the current state set, we would like to know how well the whole state set is approximated by the new one with fixed dimension. When a new state $\vx^*_t$ is observed at t, assume we can represent the kernel function as
\begin{equation}\label{k_ind}
\vK(\vX_t,\vx^*_t) = \sum_{i=1}^N \alpha_i \vK(\vX_t,\vx_{it}),
\end{equation}
where $\alpha_i$ is a coefficient, and each $\vx_{it}$ is a element in state set $\mathcal{X}$ at t. Eq. (\ref{k_ind}) means that the kernel can be represented without the newly observed point. However in general cases the equation does not hold, therefore we introduce an error measure:
\begin{equation}\label{kernel_test}
e = \|\vK(\vX_t,\vx^*_t) - \sum_{i=1}^N \alpha_i \vK(\vX_t,\vx_{it})\|^2,
\end{equation}
where $e$ is a scalar. When the error measure $e$ is within the range of a specified threshold $e_{tol}$, the new state will not be added into the state set; when $e$ is larger than the tolerance measure, the new state should be added to the state set to maintain a reasonable approximation. Eq. (\ref{kernel_test}) is called kernel independence test \cite{csato2002sparse}. In the second case, we have to delete one state vector from the state set to keep a fixed kernel size (use $\beta$ as the maximum size). We applied the sparse online Gaussian process method developed by Csat\'{o} and Opper \citep{csato2002sparse}, which efficiently approximate the KL divergence between the current GP and the GP with one data point missing. The data point corresponding to the largest KL divergence will be removed. The on-line algorithm can be summarized in \textbf{Algorithm 1}. For initialization, we use discretized Markov Decision Process (MDP) to find $z(\vX)$ by an eigensolver \citep{todorov2009eigenfunction}. 
\IncMargin{1em}
\begin{algorithm}
\SetKwData{Left}{left}\SetKwData{This}{this}\SetKwData{Up}{up}
\SetKwFunction{Union}{Union}\SetKwFunction{FindCompress}{FindCompress}
\SetKwInOut{Input}{input}\SetKwInOut{Output}{output}

\BlankLine
{Initialize GP-KL using MDP}\;

     \For{$t=1$ \KwTo $T$}{\label{forins}
        Observe a new state $\vx_t^*$, infer desirability $z(\vx_t^*)$ using GP as introduced in \ref{gpr_desirability}\;
        Compute optimal control policy $\vu(\vx^*_t)$ as in \ref{compute_u}\;

         \If(){$e>e_{tol}$}{\label{ut}
         Add $\vx^*$ to the state set $\mathcal{X}$ such that $\mathcal{X}^*=\{\mathcal{X},\vx^*\}$\\
             \lIf{\textsc{size}($\mathcal{X})>\beta$}{Delete one element from $\mathcal{X^*}$ using the method in \citep{csato2002sparse}}
}

Apply control policy $\vu(\vx^*_t)$ to the system\;

                     {\label{ut}
            }

     }

   \caption{Algorithm for on-line GP-KL}\label{algo_disjdecomp}
\end{algorithm}\DecMargin{1em}

\section{Nystr\"{o}m Approximation for KL control}

\subsection{Desirability learning via Nystr\"{o}m approximation} \label{nys_desirability}
In the last decade, the Nystr\"{o}m approximation is increasingly used as a sampling-based low-rank matrix approximator \cite{drineas2005nystrom}\citep{belabbas2009spectral}. Originally, the Nystr\"{o}m method was developed to find numerical solutions to integral equations by replacing the integral with a representative weighted sum \citep{baker1977numerical}. Suppose we have the following integral equation:
\begin{equation}
\int_a^b \vW(\vx,\vy)\phi(\vy) dy = \lambda\phi(\vx).
\end{equation}
This integral equation can be approximated by
\begin{equation}
\dfrac{b-a}{n} \sum_{j=1}^n \vW(\vx,\tilde{\vy}_j)\phi(\tilde{\vy}_j) = \lambda \phi(\vx).
\end{equation}
The approximation is based on evaluating the original integral equation at a set of evenly spaced points $\tilde{\vy}_1,\tilde{\vy}_2,\ldots,\tilde{\vy}_n$ on the interval $[a,b]$. We can solve the above equation by setting $\vx=\tilde{\vy}_i$ such that $i\in\{1,2,\ldots,n\}$. Then the equation becomes
\begin{equation}
\dfrac{b-a}{n} \sum_{j=1}^n \vW(\tilde{\vy}_i,\tilde{\vy}_j)\phi(\tilde{\vy}_j) = \lambda \phi(\tilde{\vy}_i).
\end{equation}
Here we set $[a,b]$ to be $[0,n]$ without loss of generality. Then we can rewrite the equation as
$\vA\Phi = \vA\Lambda$,
where $\vA_{ij}=\vW(\tilde{\vy}_i,\tilde{\vy}_j)$. $\Phi=[\phi_1,\phi_2,\ldots,\phi_n]$ is the eigenvector of matrix $\vA$  and $\Lambda$ is a diagonal matrix with elements $\lambda_1,\lambda_2,\ldots,\lambda_n$ the corresponding eigenvalues. Therefore, solving the original integral equation problem becomes solving an eigenvalue problem. Given a new data point $\vx^*$ that is not in the set $\{\tilde{\vy}_1,\tilde{\vy}_2,\ldots,\tilde{\vy}_n\}$, we can compute its eigenvector as:
\begin{equation}
\tilde{\phi}(\vx^*) = \frac{1}{\lambda} \sum_{j=1}^n \vW(\vx^*,\tilde{\vy}_i)\phi(\tilde{\vy}_i).
\end{equation}
where $\tilde{\phi}(x^*)$ is the approximation of $\phi(x^*)$. The Nystr\"{o}m method provides a means of approximating desirability function for newly observed state. Based on the above formula, we can efficiently solve the eigenvalue problem associated with KL control. Motivated by (\ref{Discrete-MDP}), we define
\begin{equation}
\vW(\vx^*,\vx_i)=\vQ(\vx^*)\vP(\vx^*,\vx_i),~~i=1,2,\ldots,N,
\end{equation}
then we compute the approximated desirability function
\begin{equation*}
\tilde{z}(\vx^*) = \dfrac{1}{\lambda} \sum_{i=1}^N \vW(\vx^*,\vx_i)z(\vx_i),
\end{equation*}
where $z(\vx_i)$ is the desirability function for previously observed states. The main idea of this method is to use only partial state set information to firstly obtain the desirability function by solving a simpler eigenvalue problem, and then extend the eigenvectors using complete state set information (with newly observed states). Suppose the complete state space vector is given by $[\vX,\vX^*]\in\mathbb{R}^{D\times (N+T)}$, where $\vX^*$ is a vector with all newly observed state over time: $\vX^*=[\vx^*_1,\vx^*_2,\ldots,\vx^*_T]\in\mathbb{R}^{D\times T}$, and X is a prior state knowledge. The task of computing desirability function for the whole state set becomes finding the eigenvectors for the following matrix:
\begin{equation}
\vW = \left[ \begin{array}{cc}
\vQ(\vX)\vP(\vX,\vX) & \vQ(\vX) \vP(\vX,\vX^*)  \\
\vQ(\vX^*)\vP(\vX^*,\vX) & \vQ(\vX^*)\vP(\vX^*,\vX^*) \end{array} \right].
\end{equation}
Based on the Nystr\"{o}m method, we can approximate the eigenvector of the above matrix as:
\begin{equation}
\tilde{\vz}=\left[ \begin{array}{c}
z(X)  \\
\tilde{z}(\vX^*) \end{array} \right]
=\left[ \begin{array}{c}
z(X)  \\
\vQ(\vX^*)\vP(\vX^*,\vX)z(\vX)\Lambda_{\vX}^{-1} \end{array} \right].
\end{equation}
For each newly observed state $\vx^*$, the approximated desirability function can be approximated as
\begin{equation}\label{ny_for_z} 
\tilde{z}(\vx^*)=\vQ(\vx^*)\vP(\vx^*,\vX)z(\vX)\Lambda_{\vX}^{-1}. 
\end{equation}
The matrix $\tilde{\vz}\Lambda_{\vX} \tilde{\vz}^T$ is called the \textit{Nystr\"{o}m approximation} of $\vW$. However, one assumption for applying the Nystr\"{o}m approximation is that $\vW$ should be a symmetric matrix. Although $\vP(\vX,\vX^*)=\vP(\vX^*,\vX)^T$, we observed that generally the diagonal matrices $\vQ(\vX^*)\neq \vQ(\vX)$, therefore W is not symmetric. Here we use a simple approach to compensate this issue. For a newly observed state $\vx^*$, instead of computing the cost function $q(\vx^*)$,  we compute $q(\text{mean}(\vx^*,\bar{\vX}))$ where $\bar{\vX}$ is the mean of prior state space vector \vX. Intuitively, when the newly observed state $\vx^*$ is far from $\bar{\vX}$, the Nystr\"{o}m approximation would become inaccurate. It has been shown that Nystr\"{o}m method performs poorly for points located further than a particular distance from the current manifold \citep{sonday2009coarse}. However, computing (\ref{ny_for_z}) is much more efficient than computing (\ref{gp_for_z}), since no inverse of kernel matrices need to be evaluated.


\subsection{Optimal control policy based on Nystr\"{o}m method}\label{compute_u_nys}
Given the approximated desirability function $\tilde{\vz}(\vx^*)$, we can compute the optimal control policy using the same basic formula as in (\ref{ux}): $\vu(\vx^*)=\sigma^{2}\vB(\vx^*)^T\frac{\tilde{z}_{x^*}(\vx^*)}{\tilde{z}(\vx^*)}$,
where $\tilde{z}_{\vx^*}(\vx^*)=[\frac{\partial \tilde{z}(\vx^*)}{\partial \vx^{(1)*}},\ldots,\frac{\partial \tilde{z}(\vx^*)}{\partial \vx^{(d)*}},\ldots,\frac{\partial \tilde{z}(\vx^*)}{\partial \vx^{(D)*}}]$, with each element\footnotesize{
$$
\frac{\partial z(\vx^*)}{\partial \vx^{(d)*}}=\frac{\partial \vQ(\vx^*)}{\partial \vx^{(d)*}}\vP(\vx^*,\vX)z(\vX)\Lambda_X^{-1}+\frac{\partial \vP(\vx^*,\vX)}{\partial \vx^{(d)*}}\vQ(\vx^*)z(\vX)\Lambda_X^{-1},
$$}\normalsize
%
the partial derivatives $\frac{\partial \vP(\vx^*,\vX)}{\partial \vx^{(d)*}} $  $\frac{\partial \vQ(\vx^*)}{\partial \vx^{(d)*}}$ can be computed analytically given the passive dynamics  (\ref{PassiveDynamicsProbability}) and  a differentiable cost function $q(\vx)$.

\subsection{On-line update of optimal control policy}
In this subsection we use a simple but efficient approach to on-line update of control policy. When incorporating new observed state to the state set, the size of the state set would increase. For efficient implementation, we would like to limit the size of the state set. Similar to the on-line policy update for GP-KL, we define a error measure $\epsilon$ such that when the distance between newly observed state and the mean of current state space $\Vert \vx^*-\vx\Vert\leq\epsilon$, the new state will note be added into the state space.

When the new state is added into the state space, we enforce a restriction on the number of maximum allowable size of the state set. As discussed in Section \ref{nys_desirability}, the Nystr\"{o}m method yields compromised performance when new state $\vx^*$ is far from the mean of the current state vector $\bar{\vX}$. Therefore, the criteria for deleting elements from the state set depends on the distance between $\vx^*$ and elements in \vX. We measure the Euclidean distance and remove
\begin{equation}\label{remove_ny}
\vx= \operatorname*{arg\,max}_{\vx\in\mathcal{X}} \Vert \vx^*-\vx\Vert ,
\end{equation}
where the most distant state is deleted.
The Nystr\"{o}m-KL scheme is summarized in \textbf{Algorithm 2}. Initialization details will be discussed in the next section.

\IncMargin{1em}
\begin{algorithm}
\SetKwData{Left}{left}\SetKwData{This}{this}\SetKwData{Up}{up}
\SetKwFunction{Union}{Union}\SetKwFunction{FindCompress}{FindCompress}
\SetKwInOut{Input}{input}\SetKwInOut{Output}{output}
{Initialize Nystr\"{o}m-KL} using MDP\;

     \For{$t=1$ \KwTo $T$}{\label{forins}
        Observe a new state $\vx_t^*$, compute approximated desirability $\tilde{z}(\vx_t^*)$ using Nystr\"{o}m method as introduced in \ref{nys_desirability}\;
        Compute optimal control policy $\vu(\vx^*_t)$ as in \ref{compute_u_nys}\;

         \If(){$\Vert \vx^*-\bar{\vX}\Vert>\epsilon$}{\label{ut}
         Add $\vx^*$ to the state set $\mathcal{X}$ such that $\mathcal{X}^*=\{\mathcal{X},x^*\}$\\
             \lIf{\textsc{size}($\mathcal{X})>\beta$}{Delete one element from $\mathcal{X^*}$ according to (\ref{remove_ny}) }
}

Apply control policy $\vu(\vx^*_t)$ to the system\;

                     {\label{ut}
            }

     }

   \caption{Algorithm for on-line Nystr\"{o}m-KL}\label{algo_disjdecomp}
\end{algorithm}\DecMargin{1em}

\section{Numerical Results}
In this section, we focus on two dynamical systems: inverted pendulum and car-on-a-hill. We will show  the desirability learning and on-line stochastic control performances of the proposed GP-KL and Nystr\"{o}m-KL frameworks. 

\subsection{Inverted pendulum}
The passive dynamics for the inverted pendulum is $\valpha_{pen} (\vx)=[x_v ~ \sin{(x_p)}]^{\rT}$,
where $\vx=[x_p~ x_v]^{\rT}$ and $\vB(\vx)=[0~ 1]^{\rT}$. The first task for inverted pendulum is to move at constant velocity in either direction. Therefore, the desired behavior is a limit cycle. $v_d=2.6$ is the desired velocity in both directions. The second task is to balance the inverted pendulum at $[0,0]$.
\subsection{Car-on-a-hill}
The passive dynamics for the car-on-a-hill is:
\begin{equation*}
\valpha_{car}^{\rT} (\vx) = \left[ \begin{array}{cc}
 \frac{x_v}{(1+(x_p\exp(-x_p^2/2))^2)^{\frac{1}{2}}},  \frac{-9.8\sign(x_p)}{(1+(x_p\exp(-x_p^2/2))^{-2})^{\frac{1}{2}}} \end{array}\right],
\end{equation*}
where $\vx=[x_p;x_v]$ and $\vB(\vx)=[0;1]$. The task is to be at one of the two desired state with non-zero velocities (which means it won't stay at these states). Therefore, the desired behavior is a limit cycle as well. The desired state $[p_{d1};v_{d1}]=[-2;-2],[p_{d2};v_{d2}]=[2;2]$.

\subsection{Initialization}
Firstly, we estimate the range of the state space using sampled data obtained by propagating passive dynamics, and create a uniform grid on the constrained state space. Then we evaluate the desirability $z(\vx)$ on the grid by discretized MDP (\ref{Discrete-MDP}) given cost function $q(\vx)$ and transition matrix \cite{todorov2009eigenfunction}. The transition matrix can be computed with or without knowing the system dynamics \citep{kinjo2013evaluation}. For the inverted pendulum task, the state ranges are assumed to be $x_p\in\{-\frac{\pi}{3}, \frac{\pi}{3}\}$ and $x_v\in\{-6, 6\}$. For car-on-a-hill, the assumptions are $x_p\in \{-5, 5\}$ and $x_v\in \{-6, 6\}$.  However, the controlled dynamics could fall outside this estimated state ranges, we will address this issue in \ref{performance}. 

\subsection{Desirability learning performances}
In both examples we initialize with a 20-by-20 estimated grid, which is assumed to be our prior knowledge about the state set $\mathcal {X}$ (400 states). Then we apply GP and Nystr\"{o}m methods to approximate the desirability on a 100-by-100 state space for both tasks, the 10000 states do not include any of the element in $\mathcal {X}$. The resulting $z(\vx^*)$ are shown in Fig. \ref{z_learning}. It can be seen that both GP and Nystr\"{o}m methods yield smooth and accurate approximations given limited knowledge about the state space.

An interesting problem is when the optimal behavior involves a point attractor, e.g., balancing an inverted pendulum. In \cite{todorov2009eigenfunction}, it was reported for this class of task, the parametric eigenfunction approximation methods converge to wrong solutions. Here we use the same inverted pendulum balancing example to test the proposed nonparametric methods. Results are shown in Fig. \ref{balance}.  Both methods work effectively.

\begin{figure}[!htb]
        \begin{subfigure}[b]{0.242\textwidth}
                \includegraphics[width=\textwidth]{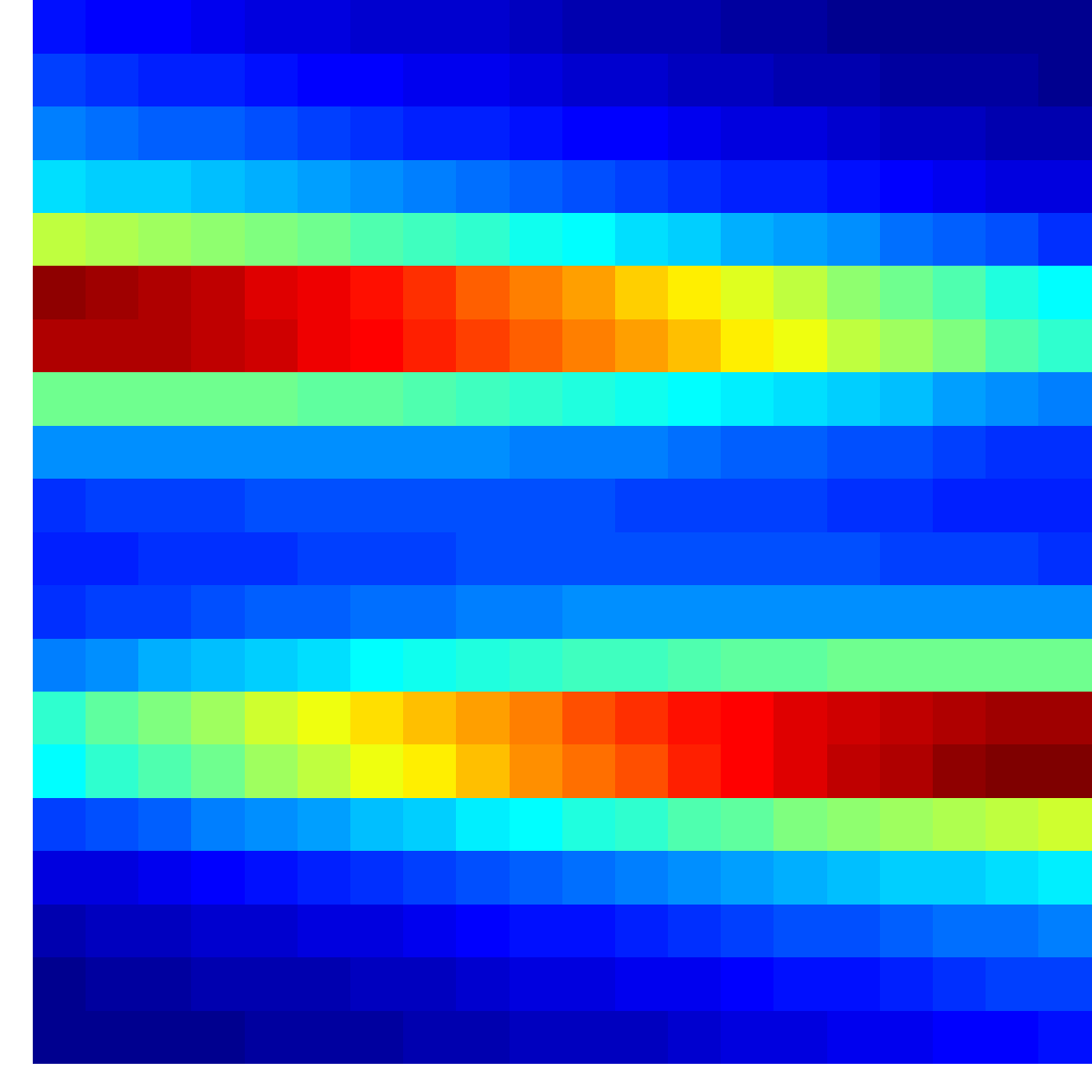}
                \caption{ }
                \label{mdp_pen}
        \end{subfigure}%
        \begin{subfigure}[b]{0.242\textwidth}
                \includegraphics[width=\textwidth]{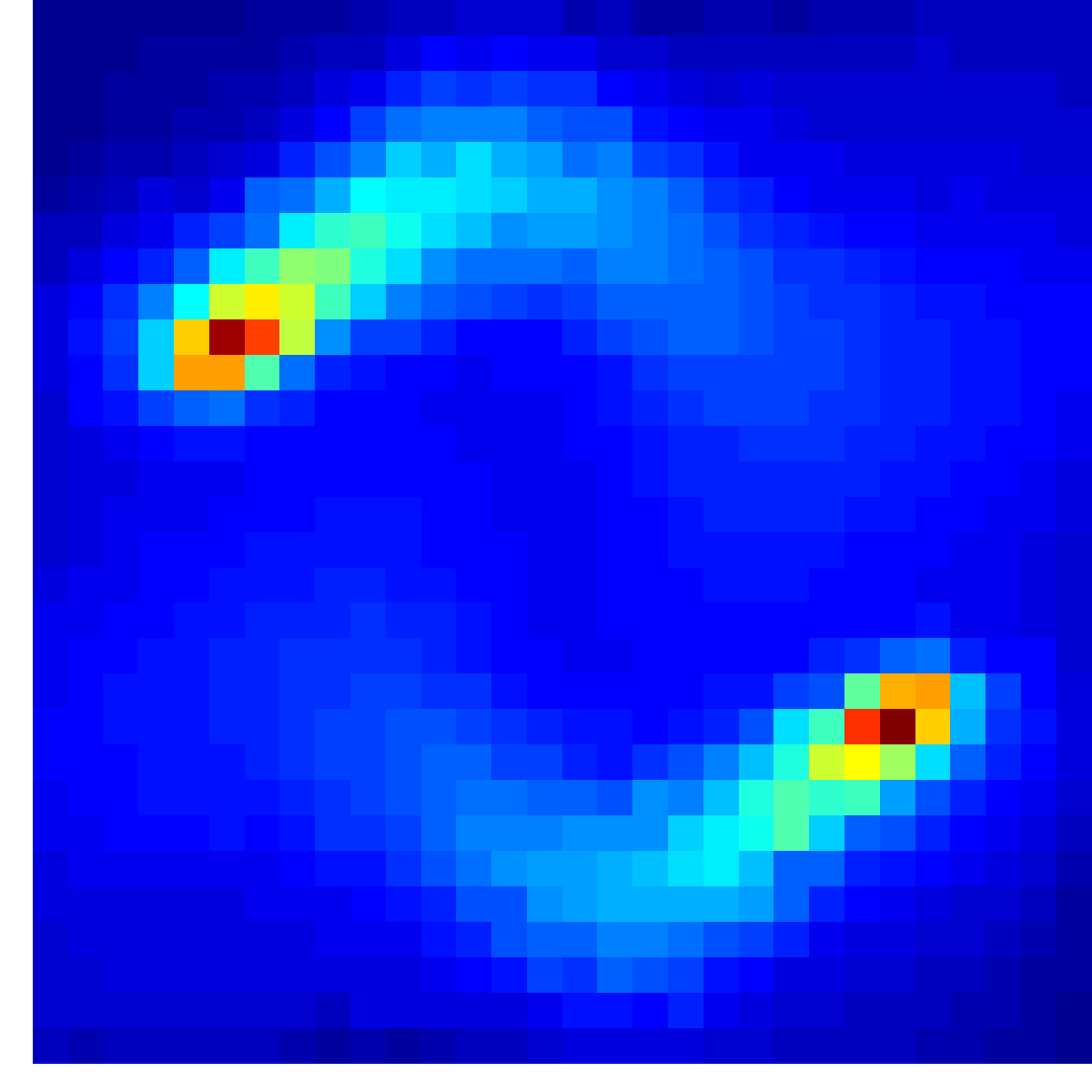}
                \caption{ }
                \label{gp_pen}
        \end{subfigure}
        \begin{subfigure}[b]{0.242\textwidth}
                \includegraphics[width=\textwidth]{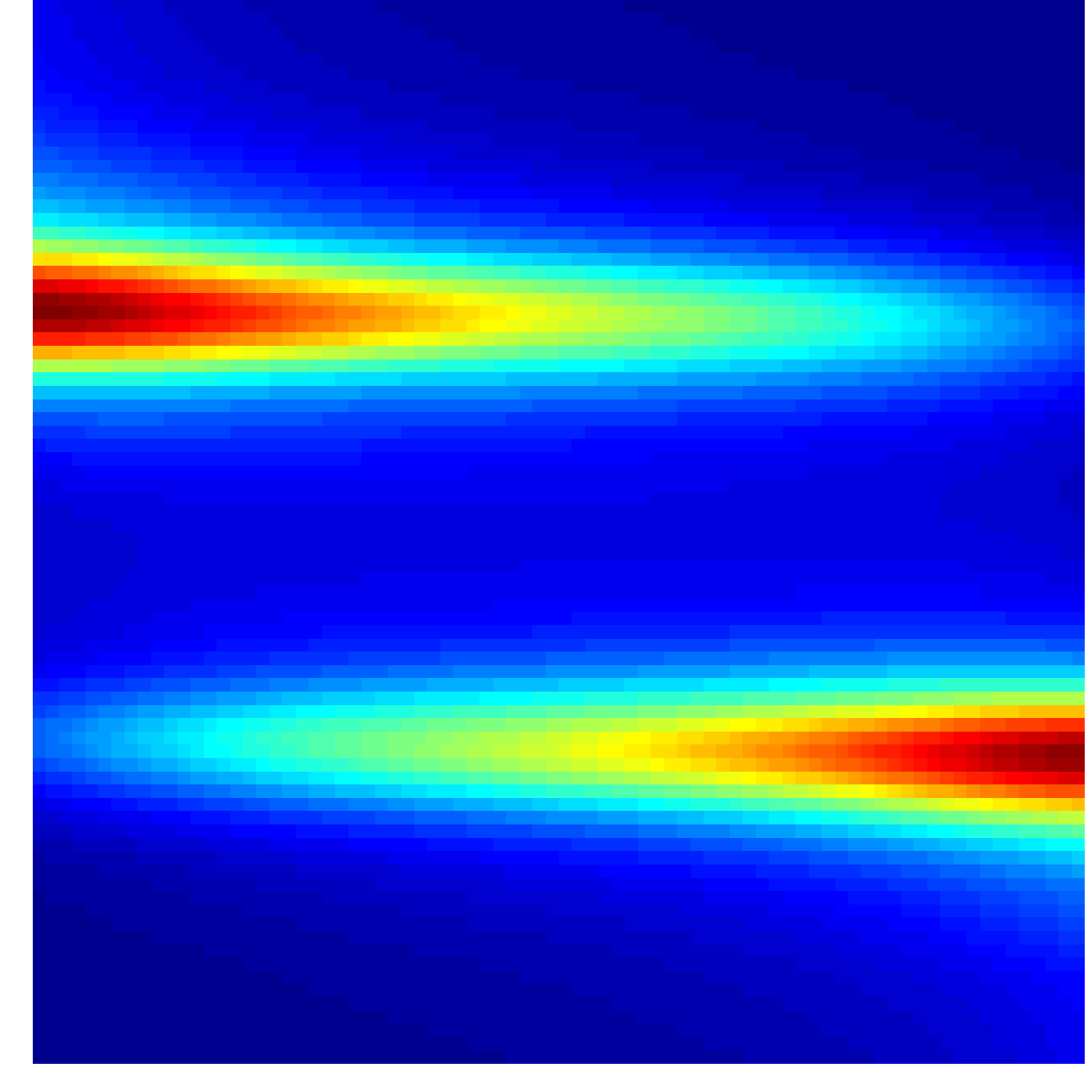}
                \caption{}
                \label{ny_pen}
        \end{subfigure}%
                \begin{subfigure}[b]{0.242\textwidth}
                \includegraphics[width=\textwidth]{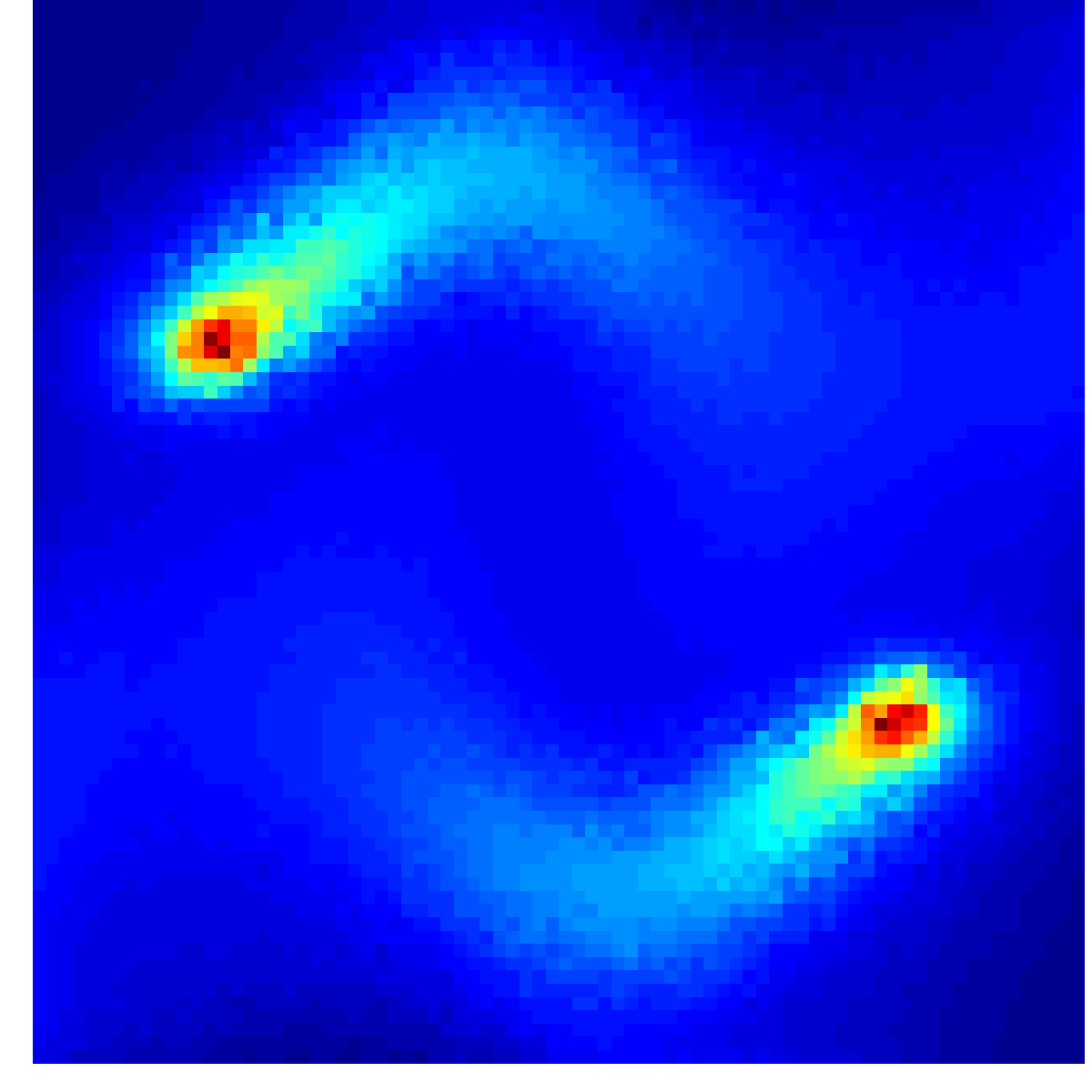}
                \caption{ }
                \label{mdp_car}
        \end{subfigure}
        \begin{subfigure}[b]{0.242\textwidth}
                \includegraphics[width=\textwidth]{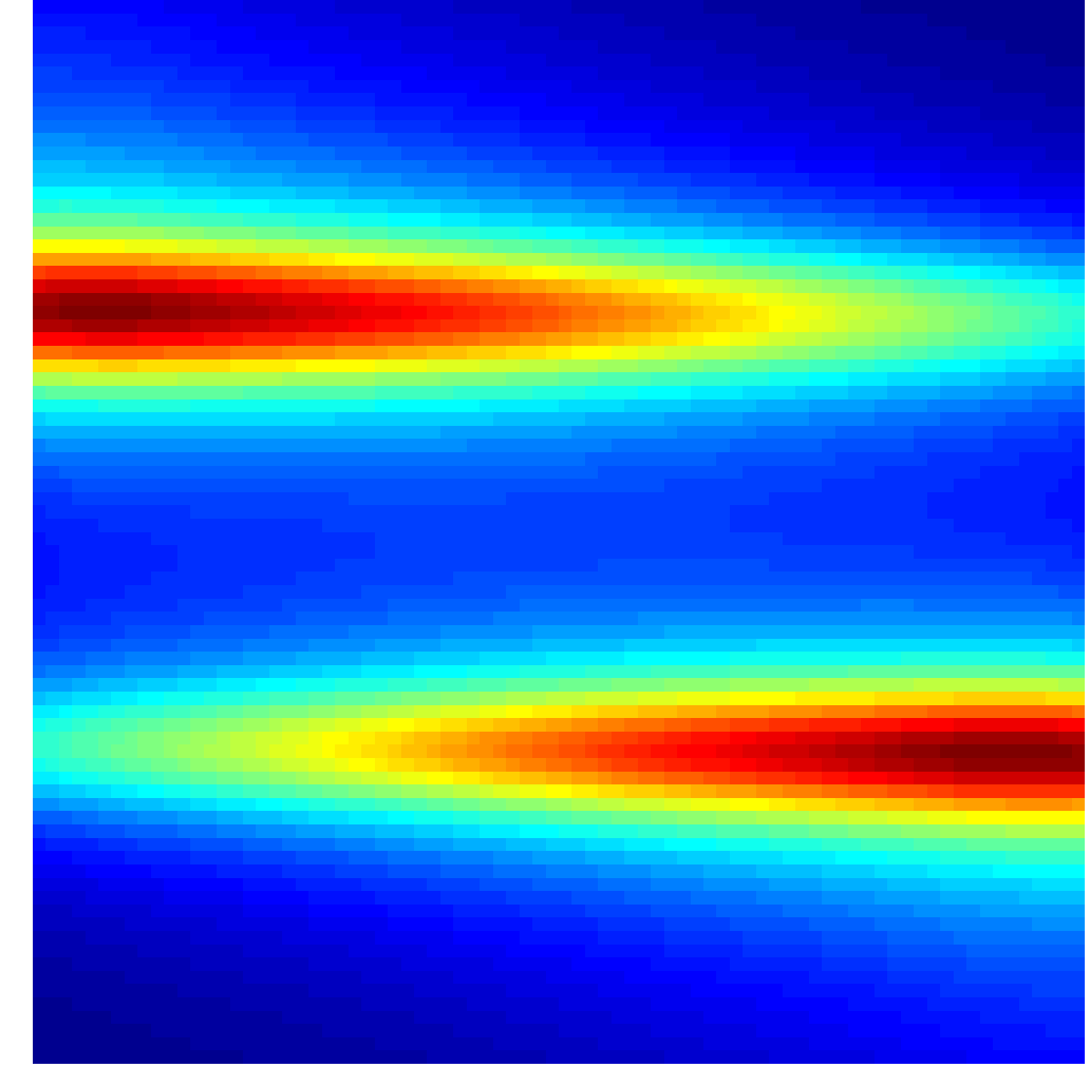}
                \caption{ }
                \label{gp_car}
        \end{subfigure}%
        \begin{subfigure}[b]{0.242\textwidth}
                \includegraphics[width=\textwidth]{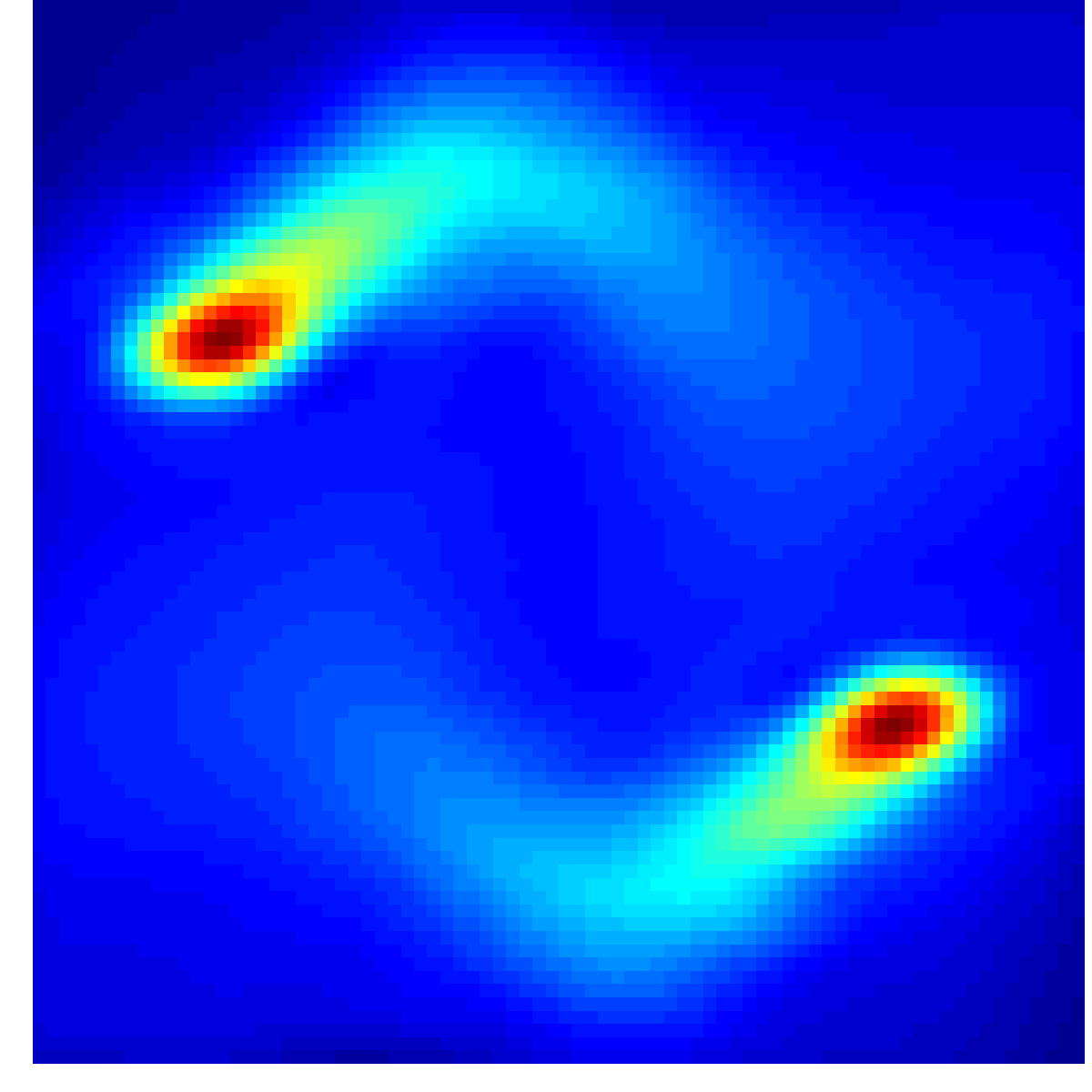}
                \caption{}
                \label{ny_car}
        \end{subfigure}
        \caption{Approximation of desirability functions. In all plots, X-axis corresponds to positions, Y-axis corresponds to velocities. Blue corresponds to smaller values and red to larger values. (a) MDP for inverted pendulum. (c) Nystr\"{o}m approximation for inverted pendulum. (e) GP for inverted pendulum. (b) MDP for - car-on-a-hill. (d) Nystr\"{o}m approximation for car-on-a-hill. (f) GP for car-on-a-hill.}\label{z_learning}
\end{figure}

\begin{figure}[!htb]
        \begin{subfigure}[b]{0.242\textwidth}
                \includegraphics[width=\textwidth]{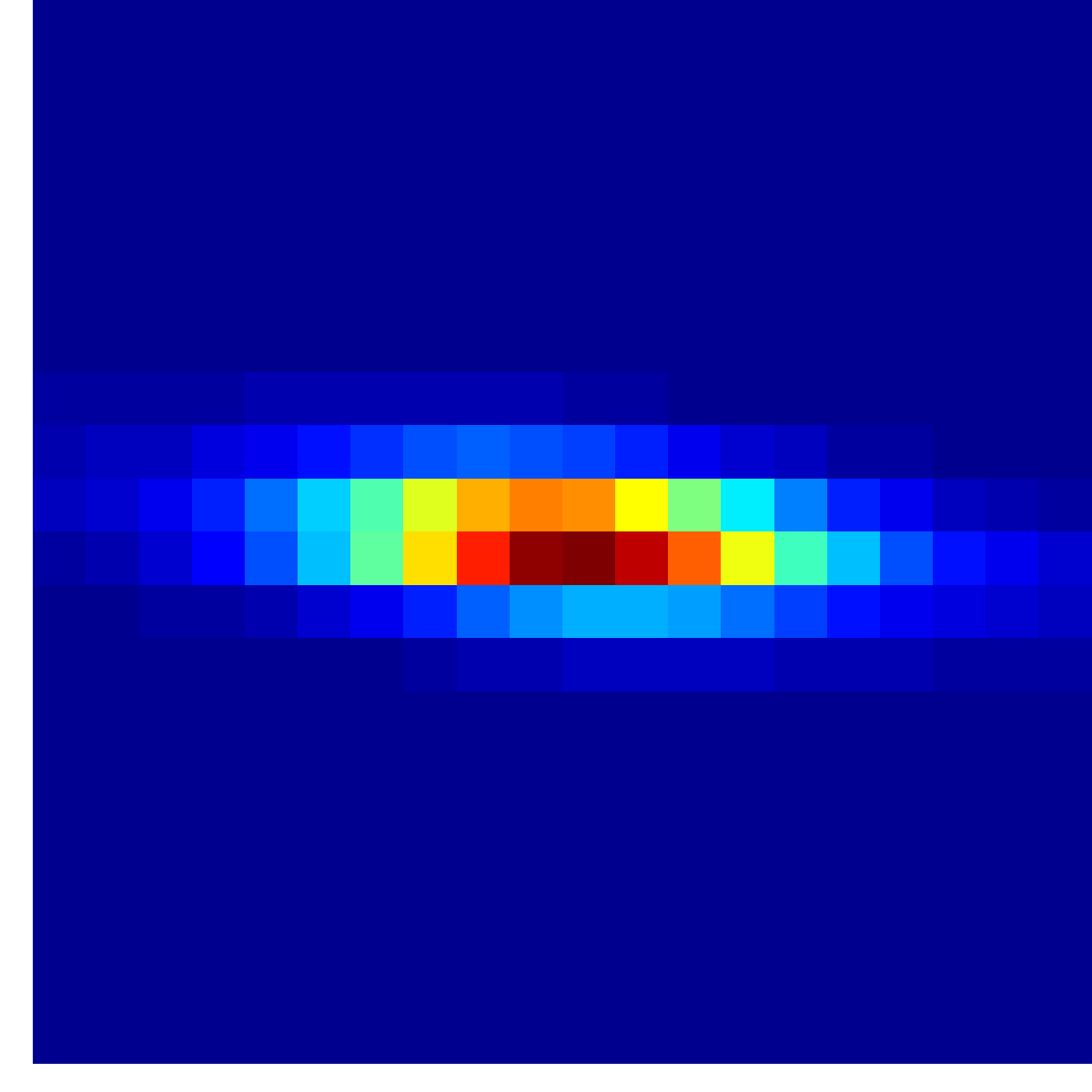}
                \caption{ }
                \label{mdp_ba}
        \end{subfigure}%
        \begin{subfigure}[b]{0.242\textwidth}
                \includegraphics[width=\textwidth]{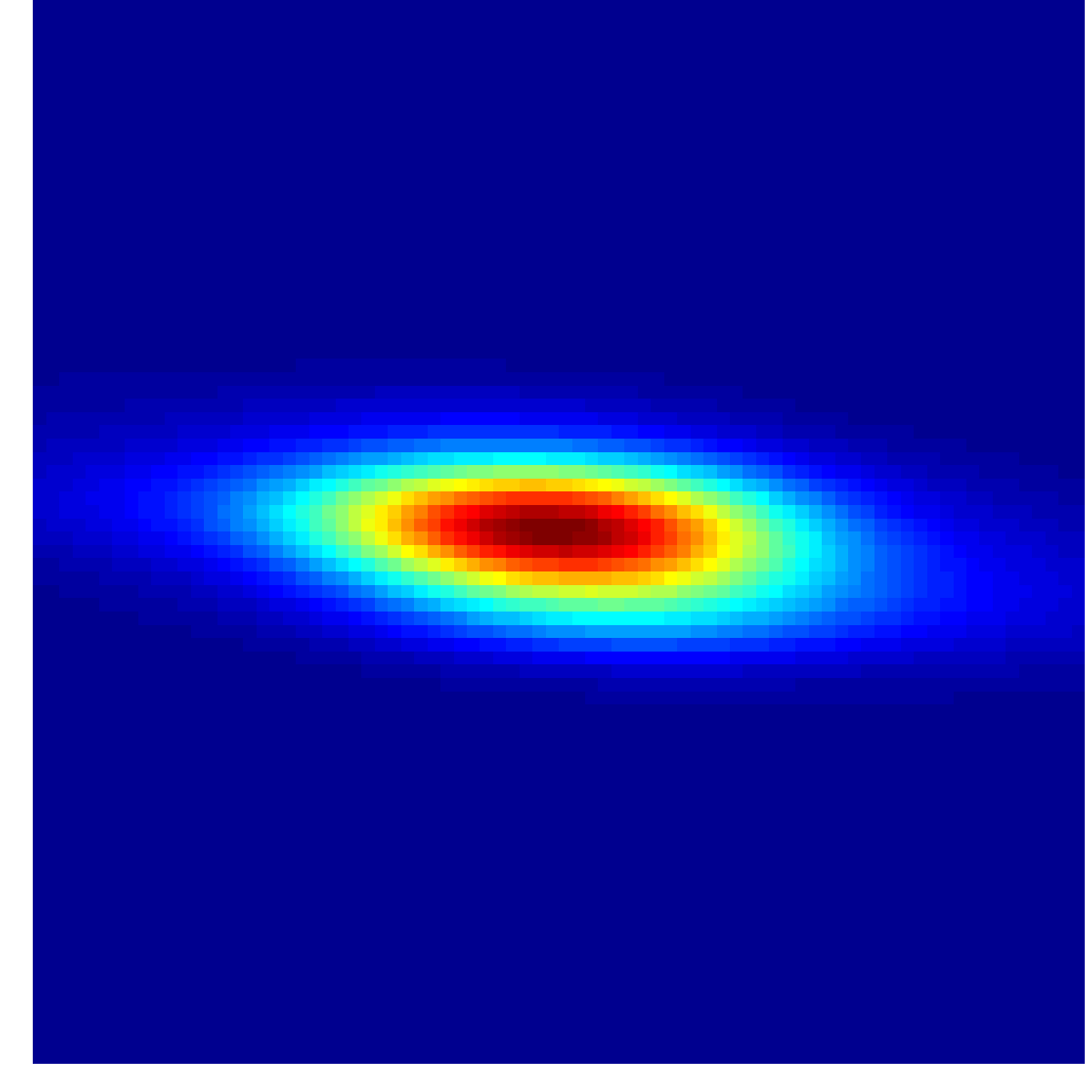}
                \caption{ }
                \label{gp_ba}
        \end{subfigure}
        \begin{subfigure}[b]{0.242\textwidth}
                \includegraphics[width=\textwidth]{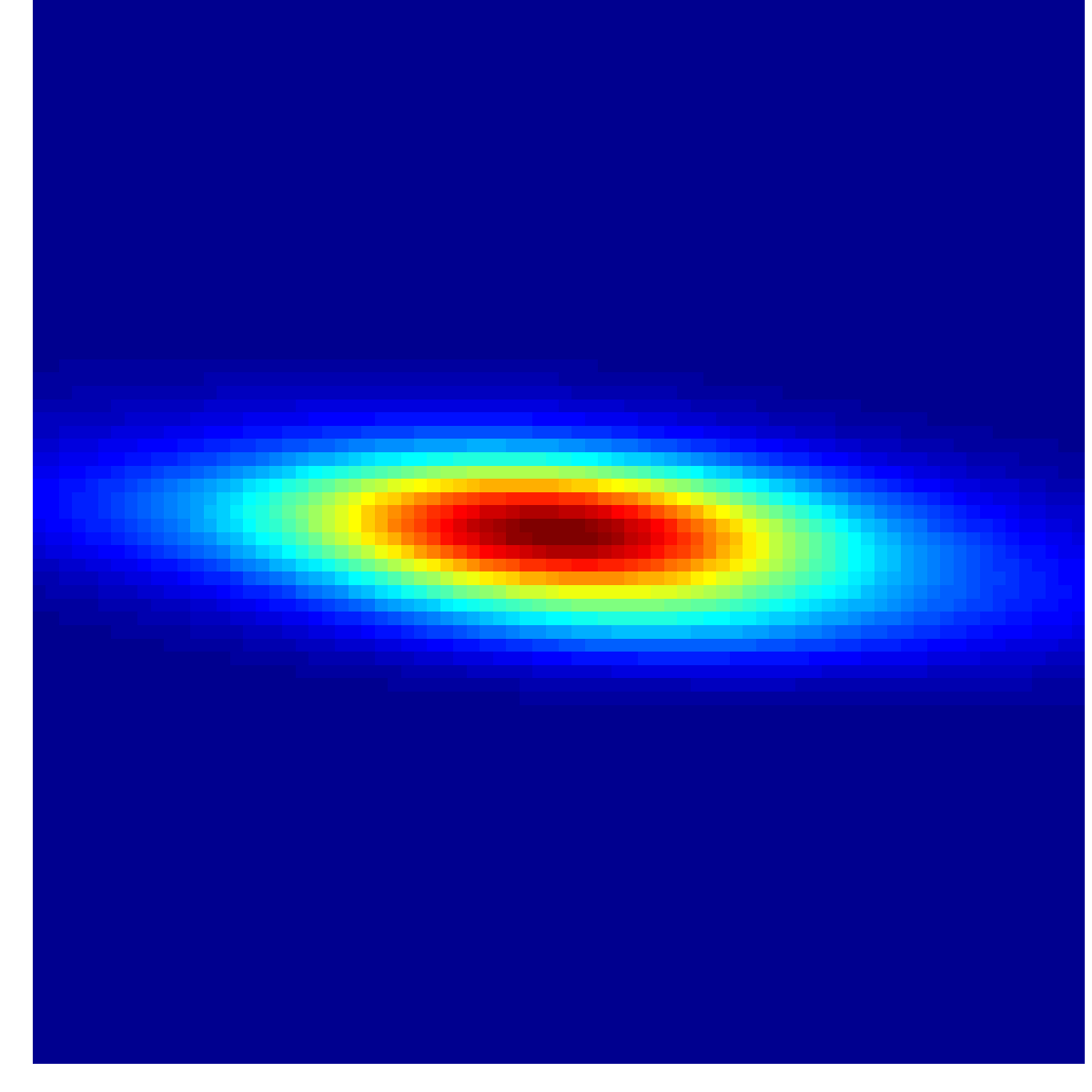}
                \caption{}
                \label{ny_ba}
        \end{subfigure}%
       \begin{subfigure}[b]{0.24\textwidth}
                \includegraphics[width=\textwidth]{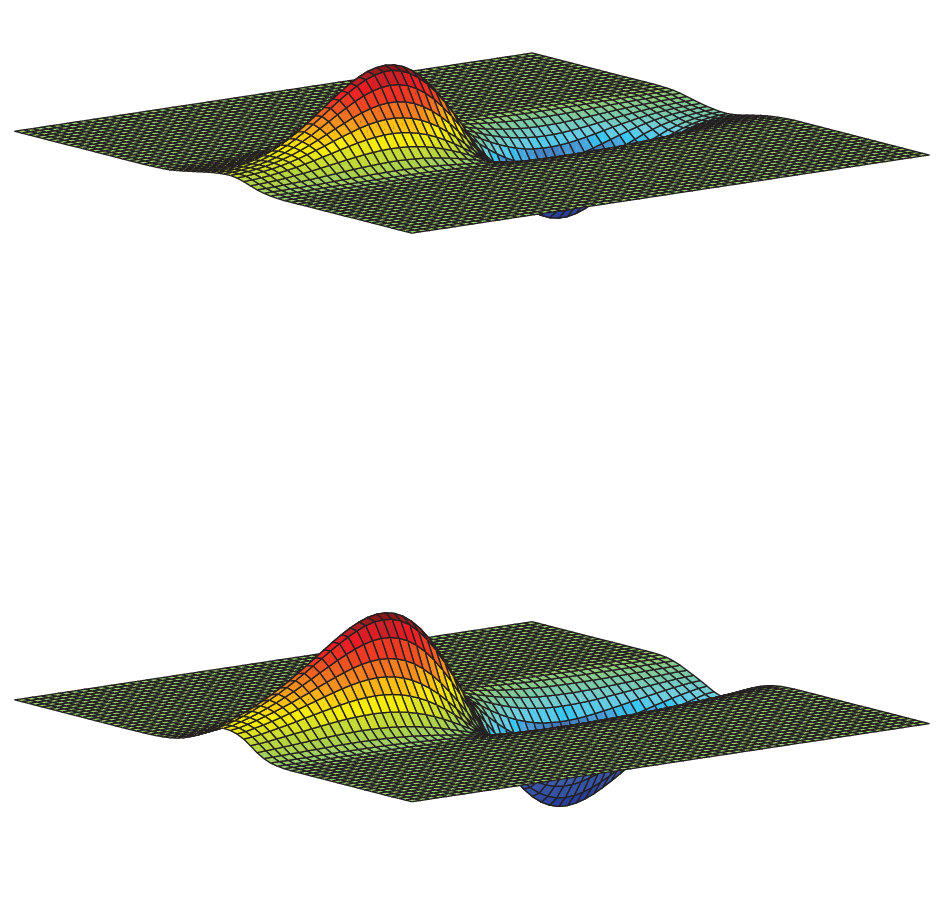}
                \caption{}
                \label{policy}
        \end{subfigure}
        \caption{ Approximation of desirability function and optimal control for inverted pendulum balancing task. (a) - MDP. (b) - Nystr\"{o}m approximation. (c) - GP. (d) - Optimal control policies on an uniformly distributed state space for GP-KL (top) and Nystr\"{o}m-KL (bottom)}\label{balance}
\end{figure}

\subsection{Control performances}\label{performance}
Now we test the on-line control performances of the proposed GP-KL and Nystr\"{o}m-KL schemes. For each task, 20 stochastic trajectories are sampled with random initial states, 500 time step per trajectory. Recall that the desired behaviors are limit cycles while keeping constant velocities or reaching specified states. Results are shown in Fig. \ref{control_performance}. The proposed on-line schemes autonomously add/remove elements in the state set and update optimal control policies according to \textbf{Algorithm 1} and \textbf{2}. As mentioned earlier, the controlled dynamics may reach the states that are far away from $\vX$ and fall outside the estimated range. For the inverted pendulum task, the actual range for position is $\{-\pi,\pi\}$ which is beyond our assumption $\{-\frac{\pi}{3}, \frac{\pi}{3}\}$. It can be seen from Fig.3 (a) and (b) that GP-KL provides slightly better performance than Nystr\"{o}m-KL. As we discussed in section \ref{nys_desirability}, the Nystr\"{o}m approach to approximating $z(\vx^*)$ yields less accurate solutions when $\vx^*$ is far away from  $\vX$. However, since the Nystr\"{o}m-KL does not compute inverse of kernel matrices as GP-KL does, it costs significantly less computational effort. While the GP-KL scheme took 71 sec, Nystr\"{o}m-KL only took 19 sec to complete the task. In the car-on-a-hill task, we assume knowing the state range (the trajectories will reach new states within the known range), both methods provide reasonable performances. The GP-KL scheme took 103 sec, and Nystr\"{o}m-KL took 32 sec. Generally speaking, GP-KL works with higher accuracy when we have inaccurate assumption of the state range. On the other hand, Nystr\"{o}m-KL demonstrates higher efficiency and is suitable when we have more confident state range assumption.

\begin{figure}[!htb]
        \begin{subfigure}[b]{0.242\textwidth}
                \includegraphics[width=\textwidth]{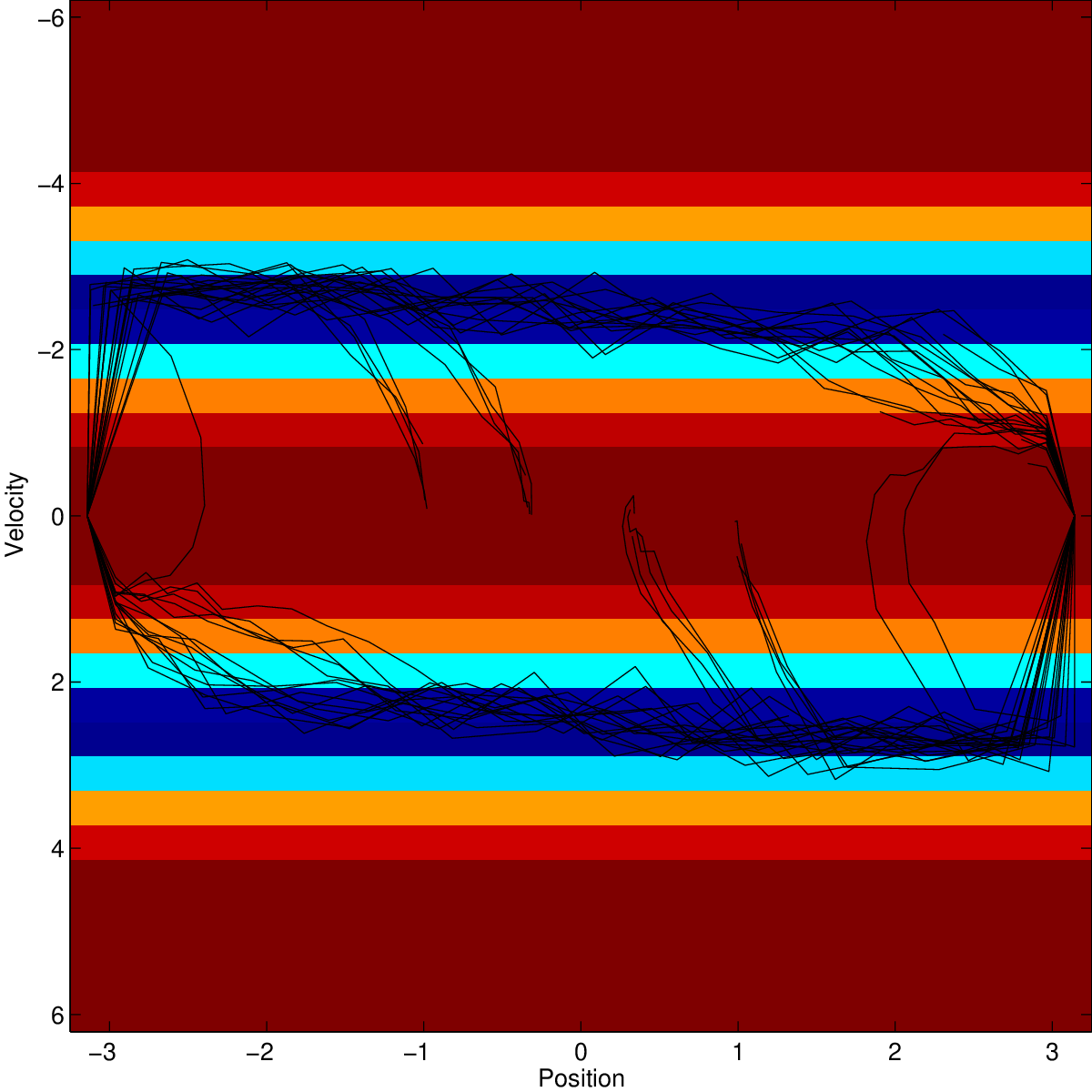}
                \caption{ }
                \label{gp_traj_pen}
        \end{subfigure}%
        \begin{subfigure}[b]{0.242\textwidth}
                \includegraphics[width=\textwidth]{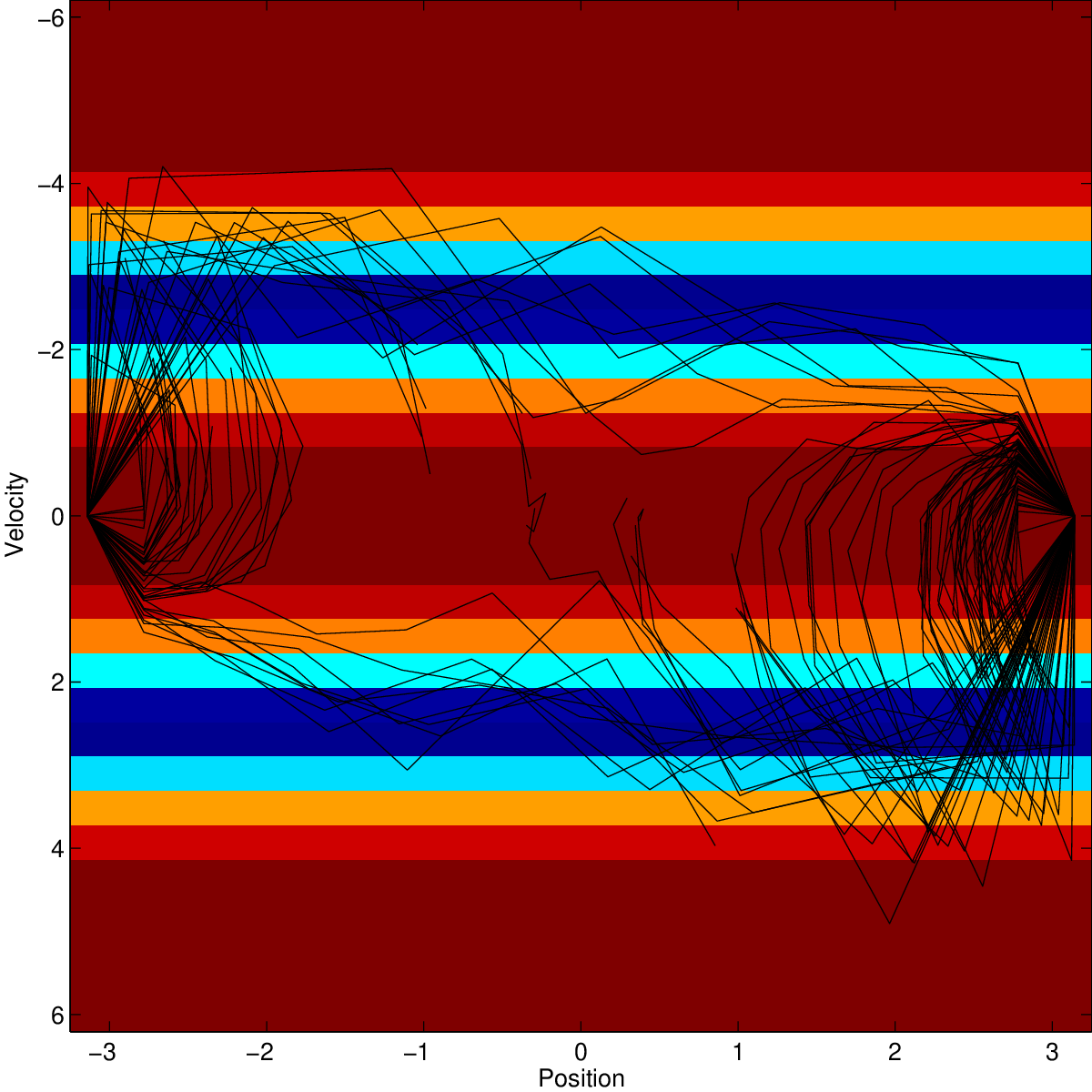}
                \caption{ }
                \label{ny_traj_pen}
        \end{subfigure}
        \begin{subfigure}[b]{0.242\textwidth}
                \includegraphics[width=\textwidth]{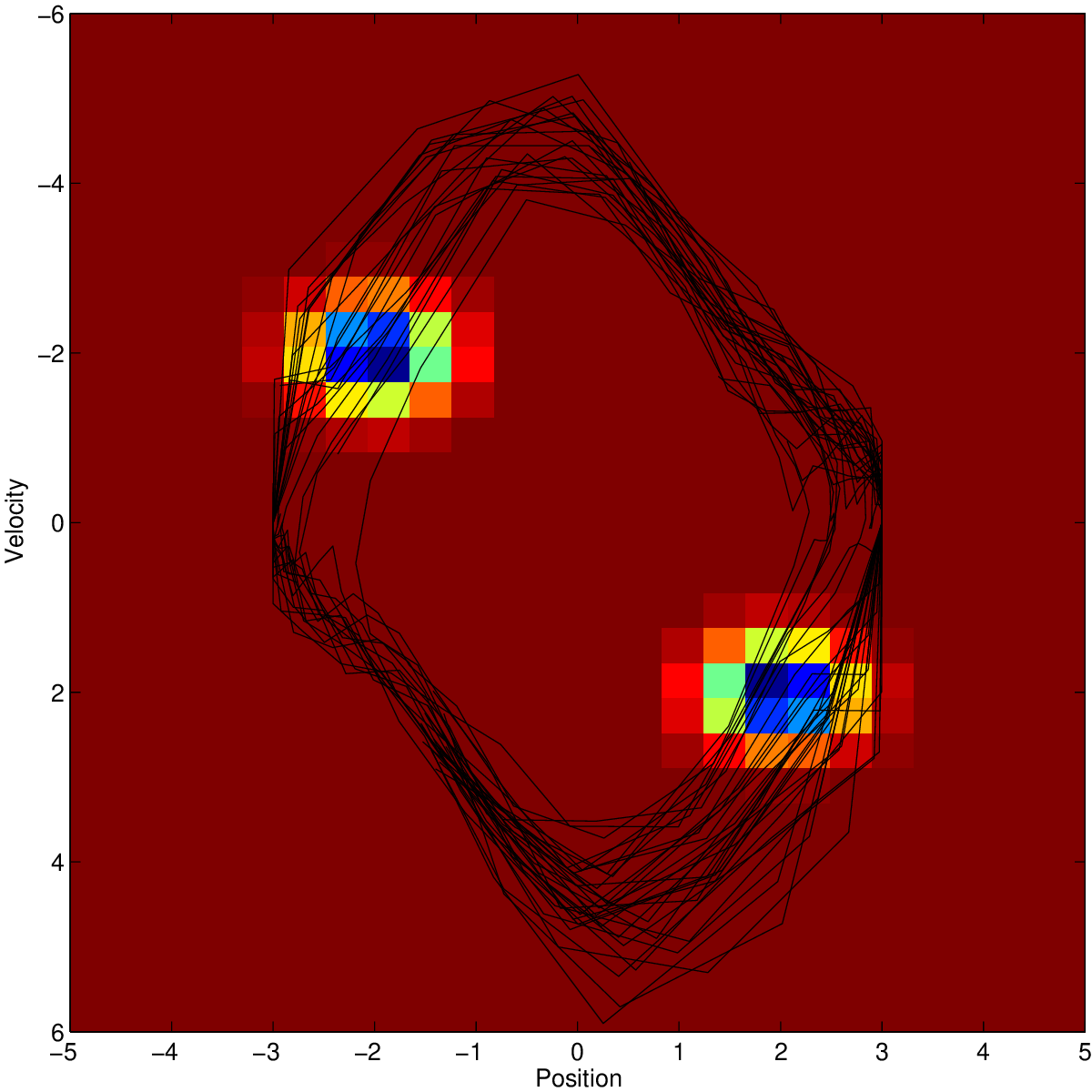}
                \caption{}
                \label{gp_traj_pen}
        \end{subfigure}%
                \begin{subfigure}[b]{0.242\textwidth}
                \includegraphics[width=\textwidth]{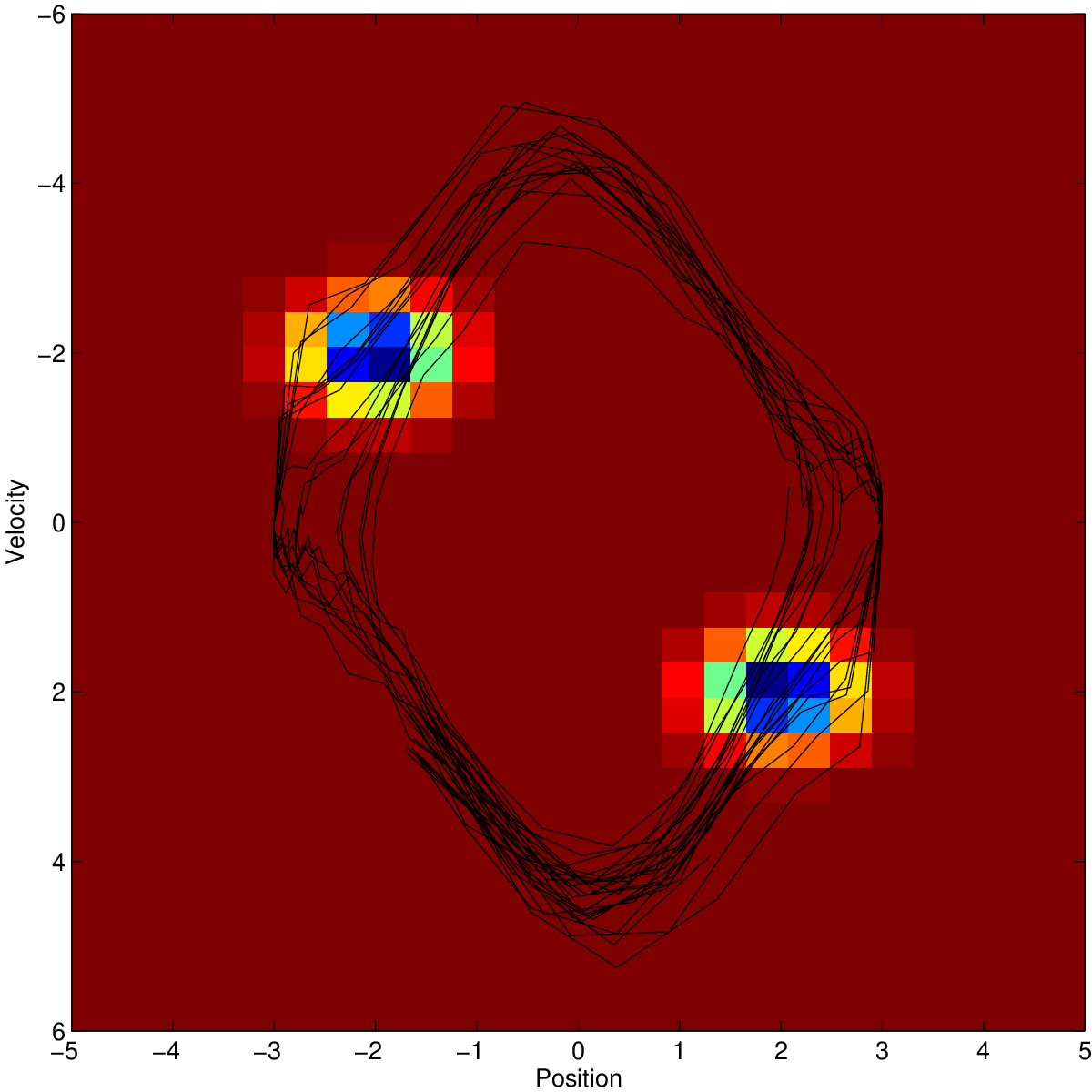}
                \caption{ }
                \label{ny_traj_car}
        \end{subfigure}
        \caption{Illustration of cost functions and controlled state trajectories. Black traces are stochastic trajectories sampled under the proposed on-line control schemes. (a) - GP-KL for inverted pendulum task. (b) - Nystr\"{o}m-KL for inverted pendulum task. (c) - GP-KL for car-on-a-hill task. (d) - Nystr\"{o}m-KL for car-on-a-hill task.}\label{control_performance}
\end{figure}

\section{Conclusions and Discussions}
Over the last decade there has been increasing number on stochastic optimal control within the machine learning community \cite{todorov2009efficient,Kappen1995,theodorou2010}  and with a plethora of applications in autonomous systems and robotics.  In this paper, we showed the mathematical connections  of KL control for infinite time horizon  problems with an   information theoretic point of view of  stochastic optimal control theory.  This view was mainly  developed within the control theory community \cite{Fleming:1995} and it is based on the relationship between free energy and relative entropy as represented by the Legendre transformation.  
  On the algorithmical side, firstly, we gave two nonparametric forms of optimal control policies based on GP and Nystr\"{o}m approximation; secondly, we proposed two frameworks for on-line update of optimal controls: GP-KL and Nystr\"{o}m-KL. Both methods feature efficient state space exploration schemes without increasing the computational demand by incorporating newly observed states and removing less necessary elements from the state training set. 
 
Compared to recently developed parametric approaches, the proposed algorithms have some notable merits: (i) accurate approximation of desirability functions; (ii) data-driven frameworks without assumed parameterization; (iii) enhanced applicability due to on-line control update with fixed computational cost.   Current limitation of the proposed approaches is that the initialization scheme requires discretization of MDP which restricts the scalability of the algorithms. The initialization scheme can be improved by various means such as local approximation of desirability function, non-uniform sampling methods to create local state grids, low-dimensional manifold embedding of high-dimensional state space, etc.
Many challenging tasks requires in-depth exploration, and our future work  will focus on further improving the applicability of both frameworks.

\bibliographystyle{unsrt}
\bibliography{references}

\begin{thebibliography}{15}
\providecommand{\natexlab}[1]{#1}
\providecommand{\url}[1]{\texttt{#1}}
\expandafter\ifx\csname urlstyle\endcsname\relax
  \providecommand{\doi}[1]{doi: #1}\else
  \providecommand{\doi}{doi: \begingroup \urlstyle{rm}\Url}\fi

\bibitem[Baker and Baker(1977)]{baker1977numerical}
C.~Baker and C.~Baker.
\newblock \emph{The numerical treatment of integral equations}, volume~13.
\newblock Clarendon press Oxford, 1977.

\bibitem[Belabbas and Wolfe(2009)]{belabbas2009spectral}
M.~Belabbas and P.~Wolfe.
\newblock Spectral methods in machine learning and new strategies for very
  large datasets.
\newblock \emph{Proceedings of the National Academy of Sciences}, 106\penalty0
  (2):\penalty0 369--374, 2009.

\bibitem[Csat{\'o} and Opper(2002)]{csato2002sparse}
L.~Csat{\'o} and M.~Opper.
\newblock Sparse on-line gaussian processes.
\newblock \emph{Neural Computation}, 14\penalty0 (3):\penalty0 641--668, 2002.

\bibitem[Drineas and Mahoney(2005)]{drineas2005nystrom}
P.~Drineas and M.~Mahoney.
\newblock On the nystr{\"o}m method for approximating a gram matrix for
  improved kernel-based learning.
\newblock \emph{The Journal of Machine Learning Research}, 6:\penalty0
  2153--2175, 2005.

\bibitem[Fleming and McEneaney(1995)]{Fleming:1995}
W.~H. Fleming and W.~M. McEneaney.
\newblock Risk-sensitive control on an infinite time horizon.
\newblock \emph{SIAM J. Control Optim.}, 33:\penalty0 1881--1915, November
  1995.

\bibitem[Kappen(2005)]{Kappen1995}
H.~J. Kappen.
\newblock Linear theory for control of nonlinear stochastic systems.
\newblock \emph{Phys Rev Lett}, 95:\penalty0 200--201, 2005.

\bibitem[Kinjo et~al.(2013)Kinjo, Uchibe, and Doya]{kinjo2013evaluation}
K.~Kinjo, E.~Uchibe, and K.~Doya.
\newblock Evaluation of linearly solvable markov decision process with dynamic
  model learning in a mobile robot navigation task.
\newblock \emph{Frontiers in neurorobotics}, 7, 2013.

\bibitem[Sonday et~al.(2009)Sonday, Haataja, and Kevrekidis]{sonday2009coarse}
B.~Sonday, M.~Haataja, and I.~Kevrekidis.
\newblock Coarse-graining the dynamics of a driven interface in the presence of
  mobile impurities: Effective description via diffusion maps.
\newblock \emph{Physical Review E}, 80\penalty0 (3):\penalty0 031102, 2009.

\bibitem[Theodorou and Todorov(2012)]{TheodorouCDC2012}
E.~Theodorou and E.~Todorov.
\newblock Relative entropy and free energy dualities: Connections to path
  integral and kl control.
\newblock In \emph{2012 IEEE 51st Annual Conference on Decision and Control
  (CDC)}, pages 1466--1473, 2012.

\bibitem[Theodorou et~al.(2010)Theodorou, Buchli, and Schaal]{theodorou2010}
E.~Theodorou, J.~Buchli, and S.~Schaal.
\newblock A generalized path integral approach to reinforcement learning.
\newblock \emph{Journal of Machine Learning Research}, \penalty0 (11):\penalty0
  3137--3181, 2010.

\bibitem[Todorov(2006)]{todorov2006linearly}
E.~Todorov.
\newblock Linearly-solvable markov decision problems.
\newblock In \emph{Advances in neural information processing systems}, pages
  1369--1376, 2006.

\bibitem[Todorov(2009{\natexlab{a}})]{todorov2009efficient}
E.~Todorov.
\newblock Efficient computation of optimal actions.
\newblock \emph{Proceedings of the national academy of sciences}, 106\penalty0
  (28):\penalty0 11478--11483, 2009{\natexlab{a}}.

\bibitem[Todorov(2009{\natexlab{b}})]{todorov2009eigenfunction}
E.~Todorov.
\newblock Eigenfunction approximation methods for linearly-solvable optimal
  control problems.
\newblock In \emph{IEEE Symposium on Adaptive Dynamic Programming and
  Reinforcement Learning, 2009 (ADPRL'09)}, pages 161--168. IEEE,
  2009{\natexlab{b}}.

\bibitem[Williams and Rasmussen(2006)]{williams2006gaussian}
C.~Williams and C.~Rasmussen.
\newblock Gaussian processes for machine learning, 2006.

\bibitem[Yang and Kushner(1991)]{Kushner1991}
J.~Yang and J.~H. Kushner.
\newblock A monte carlo method for sensitivity analysis and parametric
  optimization of nonlinear stochastic systems.
\newblock \emph{SIAM Journal in Control and Optimization}, 29\penalty0
  (5):\penalty0 1216--1249, 1991.

\end{thebibliography}

\end{document}